\documentclass[a4paper, 11pt]{article}
\usepackage{alltt}
\usepackage[T1]{fontenc}
\usepackage[utf8]{inputenc}
\usepackage[english]{babel}
\usepackage{graphics}
\usepackage[dvipsnames]{xcolor}
\usepackage{amsmath, amssymb}
\usepackage{textcomp}
\usepackage{sourcesanspro} % use source sans pro as sans serif font
\usepackage[sc]{mathpazo}
\usepackage{times}
\usepackage{doi} % automatic doi-links
\usepackage[round]{natbib} % bibliography
\usepackage{multirow} % multi rows and columns for tables
\usepackage{longtable} % tables that span several pages
\usepackage{booktabs} % nicer tables
\usepackage[title]{appendix} % better appendices
\usepackage{nameref} % reference appendices with names
\usepackage[onehalfspacing]{setspace}
\usepackage[labelfont=bf,font=small]{caption}
\usepackage{pdfpages}
\usepackage{pdflscape}

\usepackage{tikz}
\usetikzlibrary{positioning, arrows}
\tikzset{
  >=stealth',
  punkt/.style={
    rectangle,
    rounded corners,
    draw=black, very thick,
    text width=6.5em,
    minimum height=10em,
    minimum width = 10em
    text centered},
  pil/.style={
    ->,
    thick,
    shorten <=2pt,
    shorten >=2pt,}
}
\definecolor{lightg}{RGB}{230,230,230}

\usepackage{geometry}
\newcommand\longtitle{Pitfalls and Potentials in Simulation Studies}
\newcommand\shorttitle{\longtitle} % if longtitle too long, change here
\newcommand\subtitle{Questionable research practices in
comparative simulation studies allow for spurious claims of superiority
of any method}
\newcommand\longauthors{Samuel Pawel\footnote{Contributed equally.}
\footnote{Corresponding author: samuel.pawel@uzh.ch \newline
  Version March 9, 2023. Licensed under CC-BY. \newline
  Published version available at \url{https://doi.org/10.1002/bimj.202200091}} , Lucas Kook$^*$, Kelly Reeve}
\newcommand\shortauthors{S. Pawel, L. Kook, K. Reeve} % if longauthors too long, change here
\newcommand\affiliation{
  \small Epidemiology, Biostatistics and Prevention Institute  \\
  \small Center for Reproducible Science \\
  \small University of Zurich,
  Hirschengraben 84, CH-8001 Zurich
}
\title{
  \vspace{-4em}
  \textbf{\longtitle} \\
  \subtitle
}
\author{
  \textbf{\longauthors} \\
  \affiliation \\
  \small\{samuel.pawel, lucasheinrich.kook, kelly.reeve\}@uzh.ch
}
\date{} % \today} % don't forget to hard-code date when submitting to arXiv!

\usepackage{hyperref}
\hypersetup{
  bookmarksopen=true,
  breaklinks=true,
  pdftitle={\shorttitle},
  pdfauthor={\shortauthors},
  pdfsubject={},
  pdfkeywords={},
  colorlinks=true,
  linkcolor=RoyalPurple,
  anchorcolor=black,
  citecolor=MidnightBlue,
  urlcolor=BrickRed,
}

\usepackage{fancyhdr}
\usepackage{accents}
\usepackage{dsfont}

\DeclareMathOperator{\Id}{Id}
\DeclareMathOperator{\IMP}{IMP}
\DeclareMathOperator{\EPV}{EPV}
\DeclareMathOperator{\prev}{prev}
\DeclareMathOperator{\BS}{\overline{BS}}
\DeclareMathOperator{\LS}{\overline{LS}}

\DeclareMathOperator{\logit}{logit}
\DeclareMathOperator{\expit}{expit}
\DeclareMathOperator{\Var}{Var}

\DeclareMathOperator{\ND}{N}
\DeclareMathOperator{\BD}{Bernoulli}

\newcommand{\ainet}{\textsc{ainet}}
\newcommand{\ie}{{i.e.},~}
\newcommand{\cf}{{cf.}~}
\newcommand{\eg}{{e.g.},~}
\newcommand{\llpen}{\lambda\left(\alpha \sum_{j=1}^p \lvert\beta_j\rvert +
\frac{1}{2} (1-\alpha) \sum_{j=1}^p \beta_j^2 \right)}
\def \wvec {\text{\boldmath$w$}}

\newcommand{\rY}{Y}
\newcommand{\rX}{\mX}

\newcommand{\ry}{y}
\newcommand{\rx}{\xvec}

\newcommand{\pkg}[1]{\textbf{#1}}
\def \xvec {\text{\boldmath$x$}}
\def \mX {\text{\boldmath$X$}}
\newcommand{\Prob}{\mathbb{P}}
\newcommand{\Ex}{\mathbb{E}}
\newcommand{\RR}{\mathbb{R}}
\def \betavec{\text{\boldmath$\beta$}}
\newcommand{\given}{\mid}

\newcommand{\shiftparm}{\betavec}

\newcommand{\linpred}{\rx^\top\shiftparm}
\newcommand{\aipen}{\lambda\left(\alpha \sum_{j = 1}^p{w_j\lvert\beta_j\rvert} + \frac{1}{2}
  (1-\alpha) \sum_{j=1}^p{w_j\beta_j^2} \right)}
\newcommand{\eparm}{\vartheta}
\newcommand{\code}[1]{\texttt{#1}}

\begin{document}
\maketitle

% Abstract
% ======================================================================
\begin{center}
  \begin{minipage}{13cm} {\small
      \rule{\textwidth}{0.5pt} \\
      {\centering \textbf{Abstract} \\
        Comparative simulation studies are workhorse tools for benchmarking
        statistical methods. As with other empirical studies, the success of
        simulation studies hinges on the quality of their design, execution and
        reporting. If not conducted carefully and transparently, their
        conclusions may be misleading. In this paper we discuss various
        questionable research practices which may impact the validity of
        simulation studies, some of which cannot be detected or prevented by the
        current publication process in statistics journals. To illustrate our
        point, we invent a novel prediction method with no expected performance
        gain and benchmark it in a pre-registered comparative simulation study.
        We show how easy it is to make the method appear superior over
        well-established competitor methods if questionable research practices
        are employed. Finally, we provide concrete suggestions for researchers,
        reviewers and other academic stakeholders for improving the
        methodological quality of comparative simulation studies, such as
        pre-registering simulation protocols, incentivizing neutral simulation
        studies and code and data sharing. }
      \rule{\textwidth}{0.4pt}  }
\end{minipage}
\end{center}

% Introduction
% ======================================================================
\section{Introduction}
%%%%%%%%%%%%%%%%%%%%%%%%%%%%%%%%%%%%%%%%%%%%%%%%%%%%%%%%%%%%%%%%%%%%%%%%%%%%%%%%

% \begin{center}
% \begin{minipage}{12cm}
% \emph{``The first principle is that you must not fool yourself and you are
% the easiest person to fool. So you have to be very careful about that.
% After you've not fooled yourself, it's easy not to fool other scientists.''}
% \end{minipage}
% \end{center}
% \begin{flushright}
% \citet[p.~12]{Feynman1974}
% \end{flushright}

Simulation studies are to a statistician what experiments are to a scientist
\citep{Hoaglin1975}. They have become a ubiquitous tool for the evaluation of
statistical methods, mainly because simulation can be used for studying the
statistical properties of methods under conditions that would be difficult or
impossible to study theoretically. In this paper we focus on simulation studies
where the objective is to compare the performance of two or more statistical
methods (\emph{comparative simulation studies}). Such studies are needed to
ensure that previously proposed methods work as expected under various
conditions, and to identify conditions under which they fail. Moreover, evidence
from comparative simulation studies is often the only guidance available to data
analysts for choosing from the plethora of available methods
\citep{Boulesteix2013, Boulesteix2017b}. Proper design and execution of
comparative simulation studies is therefore important, and results of
methodologically flawed studies may lead to misinformed decisions in scientific
and medical practice.

\begin{figure}[!htb]
\centering
{\sf \small
\begin{tikzpicture}[thick,scale=1, every node/.style={scale=0.99}]

    % nodes
    \node [rectangle, draw, rounded corners = 0.5em, minimum height = 5.5em] (truth)
    {\footnotesize
    \begin{tabular}{l}
    \multicolumn{1}{c}{\small \textbf{Truth}} \\
     Prevalence \\
     Effect type \\
     $\dots$
     \end{tabular}};

   \node [rectangle, draw, rounded corners = 0.5em, minimum height = 5.5em] (simdat) [right = 12em of truth]
   {\footnotesize
   \begin{tabular}{l}
   \multicolumn{1}{c}{\small \textbf{Simulated Data}} \\
      Training data  \\
      Test data  \\
      \phantom{$\dots$}
    \end{tabular}};

  \node [rectangle, draw, rounded corners = 0.5em, minimum height = 5.5em] (output) [right = 10em of simdat]
  {\footnotesize
  \begin{tabular}{c}
  \multicolumn{1}{c}{\small \textbf{Predictions}}
   \end{tabular}};

 % edges
 \draw [->] (truth) -- node [above] (dgp)
 {\small
 \begin{tabular}{c}
 Data Generating \\
 Process
 \end{tabular}}
 (simdat);

 \draw [->] ([yshift = 0.5em] simdat.east) -- node [above] (analysis)
 {\small
 \begin{tabular}{c}
 Model  \\
 Fitting \end{tabular}}
 ([yshift = 0.5em] output.west);

 \draw [->] ([yshift = -0.5em] output.west) to node [below] (performance)
 {\small
 \begin{tabular}{c}
 Performance \\
 Evaluation
 \end{tabular}}
 ([yshift = -0.5em] simdat.east);

  % nodes related to edges
 \node [rectangle, draw, rounded corners = 0.5em] (methods) [above = 2em of analysis]
 {\footnotesize
 \begin{tabular}{l}
 \multicolumn{1}{c}{\small \textbf{Methods}} \\
   Logistic regression \\
   Random forest \\
   $\hdots$
\end{tabular}};

\node [rectangle, draw, rounded corners = 0.5em] (params) [above = 2em of dgp]
   {\footnotesize
   \begin{tabular}{l}
     \multicolumn{1}{c}{\small \textbf{Parameters}} \\
             Sample size \\
             Missingness \\
             $\hdots$
    \end{tabular}};

\node [rectangle, draw, rounded corners = 0.5em] (metrics) [below = 2em of performance]
   {\footnotesize
   \begin{tabular}{l}
   \multicolumn{1}{c}{\small  \textbf{Metrics}} \\
    Brier score \\
    AUC \\
    $\hdots$
    \end{tabular}};

   % QRP nodes
   \node [rectangle, draw, rounded corners = 0.5em, fill = lightg] (tuning) [left = 1.5em of methods]
       {\footnotesize
       \begin{tabular}{l}
       Selective \\
       parameter tuning / \\
       method inclusion
       \end{tabular}};

    \node [rectangle, draw, rounded corners = 0.5em, fill = lightg] (switching) [right = 2em of metrics]
       {\footnotesize \begin{tabular}{l}
       Outcome \\
       switching
       \end{tabular}};

    \node [rectangle, draw, rounded corners = 0.5em, fill = lightg] (selectreport) [above = 3.45em of truth]
       {\footnotesize \begin{tabular}{l}
       Selective \\
       reporting
       \end{tabular}};

    \node [rectangle, draw, rounded corners = 0.5em, fill = lightg] (seed) [below = 3em of dgp]
       {\footnotesize \begin{tabular}{l}
       Seed tuning
       \end{tabular}};

    \node [rectangle, draw, rounded corners = 0.5em, fill = lightg] (inclusion) [left = 2em of metrics]
       {\footnotesize \begin{tabular}{l}
       Selective handling \\
       of missing values
       \end{tabular}};

  \draw [-] (params.south) to (dgp);
  \draw [-] (methods.south) to (analysis);
  \draw [-] (metrics) to (performance);
  \draw [->] (tuning.east) to (methods.west);
  \draw [->] (switching.west) to (metrics.east);
  \draw [->] (selectreport.east) to (params.west);
  \draw [->] (selectreport.south) to (truth.north);
  \draw [->] (seed.north) to ([yshift = -1em] dgp.south);
  \draw [->] (inclusion) to (performance);

\end{tikzpicture}
}
\caption{Schematic illustration of a comparative simulation study for evaluating
performance of methods for predicting binary outcomes, such as the example study
in Section~\ref{sec:study}. Questionable research
practices (in gray) can affect all aspects of the study.}
\label{fig:diagram}
\end{figure}
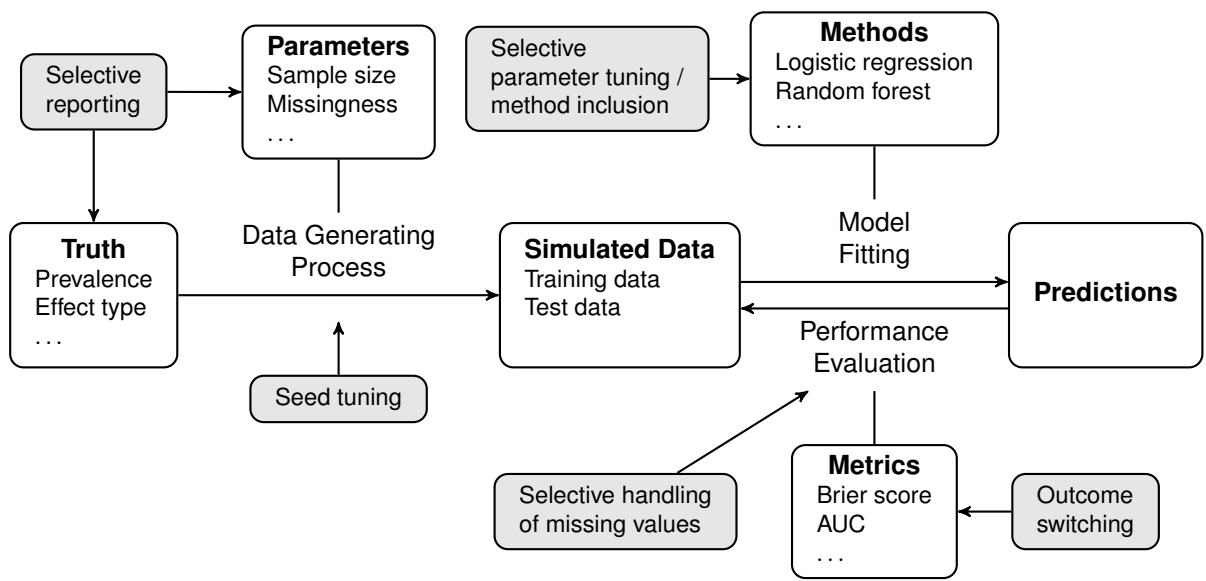

Figure~\ref{fig:diagram} shows a schematic illustration of an example
comparative simulation study. We see that, just like non-simulation based
studies, comparative simulation studies require many decisions to be made, for
instance: How will the data be generated? How often will a simulation condition
be repeated? Which statistical methods will be compared and how are their
parameters specified? How will the performance of the methods be evaluated? The
degree of flexibility, however, is much higher for simulation studies than for
non-simulation based studies as they can often be rapidly repeated under
different conditions at practically no additional cost. This is why numerous
recommendations and best practices for design, execution and reporting of
simulation studies have been proposed \citep{Hoaglin1975, Holford2000,
  Burton2006, Smith2010, OKelly2016, Monks2018, Elofsson2019, Morris2019,
  Boulesteix2020B, Chipman2022}. We recommend \citet{Morris2019} for an
introduction to state-of-the-art simulation study methodology.

%---%---%---%---%---%---%---%---%---%---%---%---%---%---%---%---%---%---%---%---
%---%---%---%---%---%---%---%---%---%---%---%---%---%---%---%---%---%---%---%---

Despite wide availability of such guidelines, statistics articles often
provide too little detail about the reported simulation studies to enable
quality assessment and replication \citep[see the literature reviews
in][]{Burton2006, Morris2019}. Journal policies sometimes require the computer
code to reproduce the results, but they rarely require or promote
rigorous simulation methodology (for instance, the preparation of a simulation protocol). This
leaves researchers with considerable flexibility in how they conduct and present
simulations studies. As a consequence, readers of statistics papers can
rarely be sure of the quality of evidence that a simulation study provides.

Unfortunately, there are many questionable research practices (QRPs) which may
undermine the validity of comparative simulations studies and which can easily
go undetected under current publishing standards. Figure~\ref{fig:diagram} shows
several QRPs that may occur in the exemplary simulation study. There is often a
fine line between QRPs and legitimate research practices. For instance, a
researcher may choose to selectively report the most relevant simulation
conditions, methods and outcomes in order to streamline the results for the
reader. These practices only become questionable when they serve to confirm the
hopes and beliefs of researchers regarding a particular method. For instance, if
only conditions and outcomes are reported where the researcher's favored method
appears superior over competitor methods. Consequently, the results and
conclusions of the study will be biased in favor of this method
\citep{Niessl2021}.

The aim of this paper is to raise awareness about the issue of QRPs in
comparative simulation studies, and to highlight the need for the adoption of
higher standards. While researchers may make decisions that can make the
conclusions of simulation studies misleading, we are not accusing them of doing
so intentionally or maliciously. Instead, we highlight how QRPs can occur and
possibly be prevented. External pressures, for example, to publish novel and
superior methods \citep{Boulesteix2015} or to concisely report large amounts of
simulation results, may also lead honest researchers to (unknowingly) employ
QRPs. As we will argue, it is not only up to the researchers but also other
academic stakeholders to improve on these issues.

This article is structured as follows: We first give an illustrative list of
QRPs related to comparative simulation studies (Section \ref{sec:QRP}). With an
exemplary simulation study, we then show how easy it is to present a novel,
made-up method as an improvement over others if QRPs are employed and \emph{a
  priori} simulation plans remain undisclosed (Section \ref{sec:study}). The
main inspiration for this work is drawn from similar illustrative studies which
have been conducted by \citet{Yousefi2009} and \citet{Jelizarow2010} for
benchmarking studies, and by \citet{Simmons2011} in the context of $p$-hacking
in psychological research. Recently, \citet{Niessl2021} and \citet{Ullmann2022}
expanded on QRPs in benchmarking studies with the latter also including
simulation studies. In Section~\ref{sec:recommendations}, we then provide
concrete suggestions for researchers, reviewers, editors and funding bodies to
alleviate the issues of QRPs and improve the methodological quality of
comparative simulation studies. Section~\ref{sec:discussion} closes with
limitations and concluding remarks.

%%%%%%%%%%%%%%%%%%%%%%%%%%%%%%%%%%%%%%%%%%%%%%%%%%%%%%%%%%%%%%%%%%%%%%%%%%%%%%%%
\section{Questionable research practices in comparative simulation studies} \label{sec:QRP}
%%%%%%%%%%%%%%%%%%%%%%%%%%%%%%%%%%%%%%%%%%%%%%%%%%%%%%%%%%%%%%%%%%%%%%%%%%%%%%%%

There are various QRPs which threaten the validity of comparative simulation
studies (see Table~\ref{table:QRPs} for an overview). QRPs can be categorized
with respect to the stage of research at which they can occur and which other
QRPs they are related to \citep{Wicherts2016}. Typically, QRPs becomes more
problematic if they are combined with related QRPs. For example, adapting the
data-generating process to achieve a desired outcome (E2) is more problematic
when the results based on the adapted process are selectively reported (R2)
compared to reporting the results based on both the original and the adapted
process. In the following, we describe QRPs from all phases of a simulation
study, namely, design, execution and reporting.

\begin{table}[!htb]
  \caption{Types of questionable research practices (QRPs) in comparative
    simulation studies at different stages of the research process. A QRP
    becomes more problematic if combined with a related QRP, especially a
    reporting QRP.}
  \label{table:QRPs}
  \centering
  \begin{tabular}{p{.05\textwidth} p{.11\textwidth} p{.75\textwidth}}
    \toprule
    \textbf{Tag} & \textbf{Related} & \textbf{Type of QRP} \\
    \midrule
    \multicolumn{2}{p{.15\textwidth}}{\textit{Design}} & \\
    D1 & E1, R1 & Not/vaguely defining objectives of simulation study \\
    D2 & E2, R1 & Not/vaguely defining data-generating process \\
    D3 & E3, E4, R1 & Not/vaguely defining which methods will be compared and how their
                      parameters are specified \\
    D4 & E1, E5, R1 & Not/vaguely defining estimands of interest \\
    D5 & E1, E5, R1 & Not/vaguely defining evaluation criteria \\
    D6 & E6, R1 & Not/vaguely defining how to handle missing values
                  (for example, due to non-convergence of methods) \\
    D7 & E7, E8, R3 & Not justifying number of simulations \\[1em]

    \multicolumn{2}{p{.15\textwidth}}{\textit{Execution}} & \\
  	E1 & D1, R2 & Changing objective of the study to achieve desired outcomes \\
    E2 & D2, R2 & Adapting data-generating process to achieve desired outcomes \\
    E3 & D3, R2 & Adding/removing comparison methods to achieve desired outcomes \\
    E4 & D3, R2 & Selective tuning of method hyperparameters to achieve desired outcomes\\
    E5 & D4, D5, R2 & Choosing evaluation criteria to achieve desired outcomes \\
    E6 & D6, R2 & Adapting inclusion/exclusion/imputation rules to achieve desired outcomes \\
    E7 & D7, R3 & Choosing number of simulations to achieve desired outcomes \\
    E8 & D7, R3 & Choosing random number generator seed to achieve desired outcomes \\[1em]

    \multicolumn{2}{p{.15\textwidth}}{\textit{Reporting}} & \\
    R1 & D1--D6 & Justifying design decisions which lead
                  to desired outcomes \emph{post hoc}\\
    R2 & E1--E6 & Selective reporting of results from simulations
                  that lead to desired outcomes \\
    R3 & D7, E7, E8 & Failing to report Monte Carlo uncertainty \\
    R4 & & Failing to assure computational reproducibility
           (for example, not sharing code and sufficient
           details about computing environment) \\
    R5 & & Failing to assure replicability (for example, not sufficiently reporting design and
           execution methodology) \\
    \bottomrule
  \end{tabular}
\end{table}

\subsection{QRPs in the design of comparative simulation studies}
The \emph{a priori} specification of research hypotheses, study design and
analytic choices is what separates \emph{confirmatory} from \emph{exploratory}
research. Evidence from confirmatory research is typically considered more
robust because study hypotheses, design, and analysis are independent of the
observed data \citep{Tukey1980}. The line between the two types of research is,
however, blurry in simulation studies since they are often iteratively
conducted, with each iteration including newly simulated data and building on
the results of the previous study. The first simulation study in a sequence of
studies may thus be exploratory whereas the subsequent studies may be
confirmatory. Yet, one may argue that in many cases a single confirmatory
simulation study which is carefully designed and whose design is justified based
on external knowledge provides more relevant evidence than a sequence of
simulation studies which are iteratively tweaked based on previous results.

To allow readers to distinguish between confirmatory and exploratory research,
many non-methodo\-logical journals require pre-registration of study design and
analysis protocols. For instance, pre-registra\-tion is common practice in
randomized controlled clinical trials \citep{DeAngelis2004}, and increasingly
adopted in experimental psychology \citep{Nosek2018} and epidemiology
\citep{Lawlor2007, Loder2010}. It is also generally recommended to write and
pre-register simulation protocols in simulation studies \citep{Morris2019}.
Well-defined study aims and methodology are arguably at least as important as in
simulation studies compared to non-simulation based studies because the space of
possible design and analysis choices is typically much larger
\citep{Hoffmann2021}. If researchers are vague or fail to define the study goals
(D1), the data-generating process (D2), the methods under investigation (D3),
the estimands of interest (D4), the evaluation metrics (D5), or how missing
values should be handled (D6) \emph{a priori} a high number of \emph{researcher
  degrees of freedom} \citep{Simmons2011} are left open. Researchers can then
generate a multiplicity of possible results which may foster overoptimistic
impressions if they report only the subset of results aligning with their hopes
and beliefs (R2), and for which they can find plausible justifications
\emph{post hoc} (R1).

Another crucial part of rigorous design is simulation size calculation
\citep[see Section~5.3 in][for an overview]{Morris2019}. While an arbitrarily
chosen, often too small, number of simulations can be executed faster, they
yield noisier results. The additional noise is not necessarily problematic if
one is only concerned with estimation. However, if the goal is to establish
method superiority through statistical tests (for instance, through a confidence
interval for the difference in method performance excluding zero), simulation
studies with too few repetitions come with undesirable properties, just as any
other study with an insufficiently large sample size. For instance, ``true''
differences in method performance are more likely to remain undetected
(increased type II errors), detected differences are more likely to be in the
wrong direction \citep[increased ``type S'' errors, see][]{Gelman2000}, and
their magnitude is more likely to be overestimated \citep[increased ``type M''
errors, see][]{Vanzwet2021}. Additionally, a researcher may start with a small
simulation size and continue to add newly simulated data until superiority is
established (\emph{optional stopping}). This is similar to early stopping of a
trial without correction for the interim analysis. Without specialized
corrections, optional stopping leads to biased estimates and increased type I
error rates \citep{Robertson2022}. These biases may also occur when the entire
simulation study is rerun with a larger sample size and the seed of the random
number generator is left unchanged. The simulated data will be the same up to
the additional data (provided the simulation runs deterministically conditional
on a seed). From this perspective, researchers should thus change the seed if
they want rerun the study and increase the simulation size adaptively.

\subsection{QRPs in the execution of comparative simulation studies}
During the execution of a simulation study researchers may (often unknowingly)
engage in various QRPs that can lead to overoptimism. For instance, the
objective of the simulation study may be changed depending on the outcome (E1).
For example, an initial comparison of predictive performance may be changed to
comparing estimation performance if the results suggest that the favored method
performs better at estimation tasks rather than prediction. The data-generating
process may also be adapted until conditions are found in which the favored
method appears superior (E2). For example, the noise levels, the number of
covariates, or the effect sizes could be changed. Competitor methods that are
superior to the proposed method may also be excluded from the comparison
altogether, or methods which perform worse under the (adapted) data-generating
process may be added (E3). The methods under comparison may come with
hyperparameters (for instance, regularization parameters in penalized regression
models). In this case, the hyperparameters of a favored method may be tuned
until the method appears superior, or the hyperparameters of competitor methods
may be tuned selectively, for example, left at their default values (E4).
Finally, the evaluation criteria for comparing the performance of the
investigated methods may also be changed to make a particular method look better
than the others (E5). For example, even though the original aim of the study may
have been to compare predictive performance among methods using the Brier score,
the evaluation criterion of the simulation study may be switched to area under
the curve if the results suggest that the favored method performs better with
respect to the latter metric. This QRP parallels the well-known
\emph{outcome-switching} problem in clinical trials \citep{Altman2017}. It is
usually not difficult to find reasonable justification for such modifications
and then present them as if they were specified during the planning of the study
(R1). As emphasized earlier, iteratively changing simulation goals, conditions,
methods under comparison and evaluation criteria can be part of finding out how
a method works. These practices become mostly problematic if only the
simulations in line with the researchers hopes and beliefs are reported (R2).

There are, however, practices which are considerably more problematic on their
own. For instance, in some simulations a method may fail to converge and thus
produce missing values in the estimates. If it is not pre-specified how these
situations will be handled, different inclusion/exclusion or imputation
strategies may be tried out until a favored method appears superior (E6).
Choosing an inadequate strategy can result in systematic bias and misleading
conclusions. If no \emph{a priori} simulation size calculation was conducted,
the simulation size may also be changed until favorable results are obtained
(E7). If in that case the number of simulations is too small, true performance
differences are more likely to be missed, their estimated direction is more
likely to be incorrect and their magnitude is more likely overestimated, as
explained previously. Finally, if only few simulations are conducted (for
instance, because the methods under investigation are computationally very
expensive), the initializing seed for generating random numbers may have a
substantial impact on the result. A particularly questionable practice in this
situation is to tune the seed until a value is found for which a preferred
method seems superior (E8).

\subsection{QRPs in the reporting of comparative simulation studies}
In the reporting stage, researchers are faced with the challenge of reporting
the design, results, and analyses of their simulation study in a digestible
manner. Various QRPs can occur at this stage. For instance, reporting may focus
on results in which the method of interest performs best (R2). Failing to
mention conditions in which the method was inferior (or at least not superior)
to competitors creates overoptimistic impressions, and may lead readers to think
that the method uniformly outperforms competitors. Similarly, presenting
simulation conditions which were added based on the observed results as
pre-planned and justified (R1) fosters overconfidence in the results.

Another crucial aspect of reporting is to adequately show the uncertainty
related to the simulation results \citep{Hoaglin1975, van2019communicating}.
Failing to report Monte Carlo uncertainty (R3), such as error bars or confidence
intervals reflecting uncertainty in the simulation, hampers the readers' ability
to assess the accuracy of the results from the simulation study and it allows
one to present random differences in performance as if they were systematic.

Finally, by failing to assure computational reproducibility of the simulation study (R4),
for example, by not sharing code and software versions to run the simulation,
it is more likely that coding errors remain undetected. By not reporting the design and
execution of the study in enough detail (R5), other researchers are unable to replicate
and expand on the simulation study.
Unclear reporting also makes it harder for readers to identify potentially overoptimistic
statements. For instance, if it is reported that all but one method are left at their
default parameters, readers can better contextualize this method's apparent superior
performance.

%%%%%%%%%%%%%%%%%%%%%%%%%%%%%%%%%%%%%%%%%%%%%%%%%%%%%%%%%%%%%%%%%%%%%%%%%%%%%%%%
\section{Empirical study: The Adaptive Importance Elastic Net
  (AINET)} \label{sec:study}
%%%%%%%%%%%%%%%%%%%%%%%%%%%%%%%%%%%%%%%%%%%%%%%%%%%%%%%%%%%%%%%%%%%%%%%%%%%%%%%%

To illustrate the application of QRPs from Table~\ref{table:QRPs} we conducted a
simulation study. The objective of the study was to evaluate the predictive
performance of a made-up regression method termed the \emph{adaptive importance
  elastic net} (\ainet). The main idea of \ainet{} is to use variable importance
measures from a random forest for a weighted penalization of the variables in an
elastic net regression model. The hope is that this \emph{ad hoc} modification
of the elastic net model improves predictive performance in clinical prediction
modeling settings where penalized regression models are frequently used.
Superficially, \ainet{} may seem sensible, however, for the data-generating
process considered in our simulation study no advantage over the classical
elastic net is expected. For more details on the method, we refer the reader to
the simulation protocol (Appendix~\ref{appendix:protocol}). We report the
pre-registered\footnote{We use the term \emph{pre-registered} throughout to
  refer to simulation analyses conducted as pre-specified in the protocol.}
simulation study results in the online supplement. As expected, the performance
of \ainet{} was virtually identical to standard elastic net regression. \ainet{}
also did not yield any improvements over logistic regression for the
data-generating process that we considered sensible \emph{a priori} (that is,
specified based on typical conditions in clinical prediction modeling and
simulation studies from other researchers).

%--%--%--%--%--%--%--%--%--%--%--%--%--%--%--%--%--%--%--%--%--%--%--%--%--%--
\begin{figure}[!htb]
  \centering
  \includegraphics[width = 0.95\textwidth]{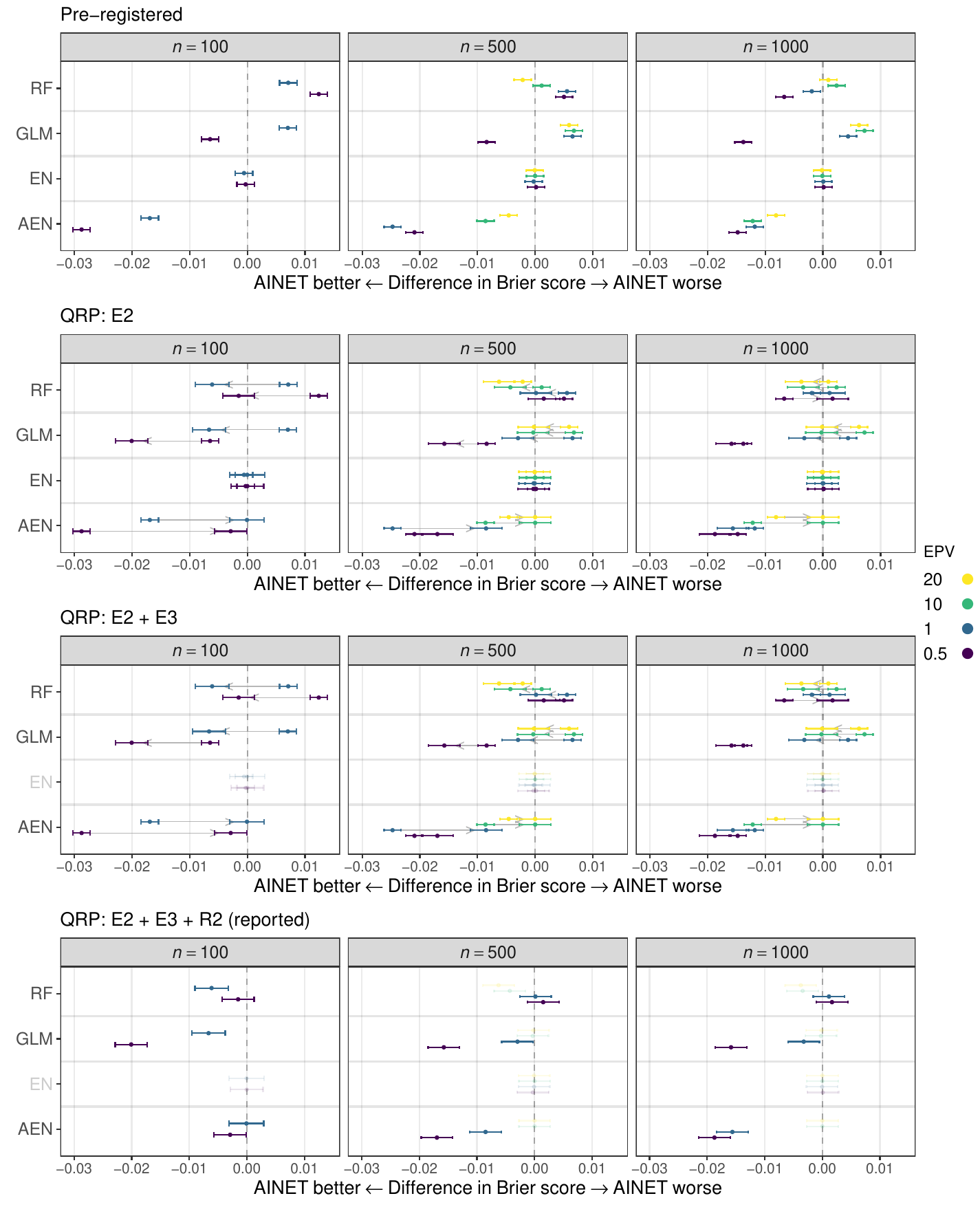}
  \caption{Differences in Brier score with 95\% adjusted confidence intervals
    between \ainet{} and random forest (RF), logistic regression (GLM), elastic
    net (EN) and adaptive elastic net (AEN) are shown for representative
    simulation conditions (correlated covariates $\rho = 0.95$, prevalence
    $\operatorname{prev} = 0.05$, a range of sample sizes $n$ and events per
    variable (EPV), in each simulation the Brier score is computed for 10'000
    test observations; for details see Appendix~\ref{appendix:protocol}). The
    top row depicts the pre-registered results in which \ainet{} does not
    outperform any competitor uniformly, except AEN. In the second row, we apply
    QRP E2: Altering the data-generating process by adding a non-linear effect
    and sparsity. The gray arrows point from the pre-registered result to the
    results under the tweaked simulation. In the third row, QRP E3 is applied:
    EN is removed as a competitor. In the bottom row, selective reporting R2 is
    applied: Only low EPV settings are reported to give a more favorable
    impression for \ainet{}. Arrows are depicted only for non-overlapping
    confidence intervals. } \label{fig:E1}
\end{figure}
%--%--%--%--%--%--%--%--%--%--%--%--%--%--%--%--%--%--%--%--%--%--%--%--%--%--

We now show how application of QRPs changes the above pre-registered
conclusions. Figure~\ref{fig:E1} illustrates different types of QRPs
sequentially applied to simulation-based evaluation of \ainet{}. The top row
depicts the pre-registered differences in Brier score (horizontal axis) between
\ainet{} and competitor methods (vertical axis) for a representative subset of
the simulation conditions. A negative difference indicates superior performance
of \ainet{}. In the second row, the arrows depict the change in the
pre-registered results after changing the data-generating process (E2). The
third row shows the result after removal of the elastic net competitor (E3).
Finally, the bottom row shows the end result where selective reporting of
simulation conditions and competitor methods (R2) is applied to give a more
favorable impression of \ainet{}. We will now discuss these QRPs in more detail.

\paragraph{Altering the data-generating process (E2)}
We could not detect a systematic performance benefit of \ainet{} over standard
logistic regression, elastic net regression, or random forest for the scenarios
specified in the protocol. For this reason, we tweaked the data-generating
process by adding different sparsity conditions and a non-linear effect. We then
found that \ainet{} outperforms logistic regression under the following
conditions: Only few variables being associated with the outcome (sparsity), a
non-linear effect and a low number of events per variable (EPV).
Figure~\ref{fig:E1} (second row) shows the changes in Brier score difference
between the pre-registered and the tweaked simulation. As can be seen, the
tweaked data-generating process leads to \ainet{} being superior to competitors
in some conditions, and at least not inferior in others.

\paragraph{Removing competitor methods (E3)}
Despite the adapted data-generating process, we still observed only minor (if
any) improvements of \ainet{} over the elastic net. In order to present \ainet{}
in a better light we could omit the comparisons with the elastic net (E3), as
shown in Figure~\ref{fig:E1} (third row). This could be justified, for example,
by arguing that for neutral comparison it is sufficient to compare a less
flexible method (logistic regression, which has no tuning parameters and
captures linear effects), a more flexible method (random forest, which has
tuning parameters and captures nonlinear relationships), and a comparably
flexible method (adaptive elastic net, which has the same tuning parameters as
AINET, but differs in the way the penalization weights are chosen).

\paragraph{Selective reporting of simulation results (R2)}
After the removal of the competitor elastic net, there are still some simulation
conditions under which \ainet{} is not superior to the remaining competitors. To
make \ainet{} appear more favorable, we thus report only simulation conditions
with low EPV, as shown in Figure~\ref{fig:E1} (fourth row). This could be
justified by the fact that journals require authors to be concise in their
reporting. Otherwise, further conditions with low EPV values could be simulated
to make the results seem more exhaustive. Focusing primarily on low EPV settings
could be justified in hindsight by framing \ainet{} as a method designed for
high-dimensional data (low sample size relative to the number of variables).

%%%%%%%%%%%%%%%%%%%%%%%%%%%%%%%%%%%%%%%%%%%%%%%%%%%%%%%%%%%%%%%%%%%%%%%%%%%%%%%%
\section{Recommendations}
\label{sec:recommendations}
%%%%%%%%%%%%%%%%%%%%%%%%%%%%%%%%%%%%%%%%%%%%%%%%%%%%%%%%%%%%%%%%%%%%%%%%%%%%%%%%

The previous sections painted a rather negative picture of how undisclosed
changes in simulation design, analysis and reporting may lead to overoptimistic
conclusions. In the following, we summarize what we consider to be practical
recommendations for improving the methodological quality of simulation studies;
see Table~\ref{table:recommendations} for an overview. Our recommendations are
grouped with regards to which stakeholder they concern.

\begin{table}[!htb]
  \caption{Recommendations for improving quality of comparative simulation studies
    and preventing QRPs.}
  \label{table:recommendations}
  \centering
  \begin{tabular}{p{.95\textwidth}}
    \toprule
    \textit{Researchers} \\
    \midrule
    \vspace{-1.5em}
    \begin{itemize}
      \setlength\itemsep{0pt}
      \setlength\itemindent{-12pt}
      \item[--] Write (and possibly pre-register) simulation protocols
      \item[--] Adopt good computational practices (code review, packaging, unit-tests)
      \item[--] Share code and data (possibly in intermediate/summary form to enable
      secondary analysis)
      \item[--] Report the process of the simulation study fully and transparently
      (for instance, time-stamped protocol amendments to disclose pilot studies
      and \emph{post hoc} modifications)
      \item[--] Perform simulation analysis in a blinded manner
      \item[--] Collaborate with other research groups (possibly familiar with ``competing'' methods)
      \item[--] Disclose multiplicity and uncertainty of results (for example, with sensitivity analyses)
      \item[--] Teach simulation study methodology in statistics (post)graduate courses
    \end{itemize}

    \textit{Editors and reviewers} \\
    \midrule
    \vspace{-1.5em}
    \begin{itemize}
      \setlength\itemsep{0pt}
      \setlength\itemindent{-12pt}
      \item[--] Encourage exploration of conditions where methods should
      be inferior or break down
      \item[--] Encourage (pre-registered) simulation protocols
      \item[--] Provide enough space for description of simulation methodology
    \end{itemize}

    \textit{Journals and funding bodies} \\
    \midrule
    \vspace{-1.5em}
    \begin{itemize}
      \setlength\itemsep{0pt}
      \setlength\itemindent{-12pt}
      \item[--] Provide incentives for rigorous simulation methodology (such as badges on papers)
      \item[--] Require code and data
      \item[--] Promote standardized reporting
      \item[--] Adopt reproducibility checks
      \item[--] Promote/fund research and software to improve simulation study methodology
      \item[--] Shift focus away from outperforming state-of-the-art methods
    \end{itemize}\\
    \bottomrule
  \end{tabular}
\end{table}

\subsection{Recommendations for researchers}
Adopting pre-registered simulation protocols is an important measure that
researchers can take to prevent themselves from subconsciously engaging in QRPs.
Pre-registration enables readers to distinguish between confirmatory and
exploratory findings, and it lowers the risk of potentially flawed methods being
promoted as an improvement over competitors. While pre-registered simulation
protocols may at first seem disadvantageous due to the additional work and
possibly lower chance of publication, they provide researchers with the means to
differentiate their high-quality simulation studies from the numerous
unregistered and possibly less trustworthy simulation studies in the literature.
Platforms such as GitHub (\url{https://github.com/}), OSF
(\url{https://osf.io/}), or Zenodo (\url{https://zenodo.org/}) can be used for
archiving and time-stamping documents. Moreover, pre-registration can also save
researchers from some work later on. For instance, large parts of the
methodology description can usually be copied from the protocol to the final
manuscript.

When pre-registering and conducting simulation studies, we recommend using a
robust computational workflow. Such a workflow encompasses packaging the
software, writing unit tests and reviewing code \citep[see][]{Schwab2021}. Other
researchers and the authors themselves then benefit from improved computational
reproducibility and less error-prone code. Of course, there are also certain
practical limits to computational reproducibility. For instance, if a simulation
study requires high performance computing and/or several weeks of running time,
the authors should not expect reviewers and journals to replicate their
simulation study from scratch. The authors should nevertheless provide the code
to run the simulation and, if possible, they should also provide intermediate
simulation results (for instance, fitted model objects) so that the simulation
study can at least be partially reproduced. Similarly, authors can share the
simulated data, either in raw and/or some summarized form (for example, sharing
simulated data sets and parameter estimates of fitted models). This allows
interested readers and reviewers to do additional analyses. Unlike experiments
with human subjects, there are no privacy concerns for sharing simulation data.
Furthermore, online tools, such as INTEREST \citep[INteractive Tool for
Exploring REsults from Simulation sTudies,][]{Gasparini2021}, can be used for
interactive exploration of the data set.

While planning a simulation study, it is impossible to think of all potential
weaknesses or problems that may arise when conducting the planned simulations.
In turn, researchers may be reluctant to tie their hands in a pre-registered
protocol. However, a transparently conducted and reported preliminary simulation
can obviate most of these problems. We recommend researchers to disclose
preliminary results and any resulting changes to the protocol, for example, in a
revised and time-stamped version of the protocol. This approach is similar to
conducting a small pilot study, as is often done in non-simulation based
research. Even if researchers realize that further changes are required after
the main simulation study has begun, transparent reporting of when and why
\emph{post hoc} modifications were made allows the reader to better assess the
quality of evidence provided by the study. Researchers designing simulation
studies may draw inspiration from clinical trials by tracking their protocol
modifications and time-stamping versions of their protocol.

A different approach for making \emph{post hoc} changes to the protocol is to
use blinding in the analysis of the simulation results \citep{Dutilh2019}.
Blinded analysis is a standard procedure in particle physics to prevent data
analysts from biasing their result towards their own beliefs \citep{Klein2005},
and it lends legitimacy to \emph{post hoc} modifications of the simulation
study. For instance, researchers might shuffle the method labels and only
unblind themselves after the necessary analysis pipelines are set in place. An
alternative blinding approach is to carry out data generation and analysis by
different researchers. For instance, the study from \citet{Kreutz2020} involved
two independent research groups, one who simulated and one who analyzed the
data. A related way for improving simulation studies is to collaborate with
other researchers, possibly ones familiar with ``competing'' methods. This helps
to design simulation studies which are more objective and whose results are more
useful for making a decision about which method to choose under which
circumstances.

We also recommend researchers to disclose the multiplicity and uncertainty
inherent to the design and analysis of their simulation studies
\citep{Hoffmann2021}. For instance, researchers can report sensitivity analyses
that show how the study results change for different analysis decisions (for
example, Table~4 in \citet{vanSmeden2016} shows how the evaluation metrics for
different estimators change depending on how convergence of a method is
defined). Methods from multivariate statistics can be used for visualizing the
influence of different design choices, such as the multidimensional unfolding
approach in \citet{Niessl2021}.

One reason for the low standards of simulation studies in the statistics
literature may be that rigorous simulation methodology is usually not taught in
graduate or postgraduate courses (with a few exceptions, such as the course
``Using simulation studies to evaluate statistical methods'' from the MRC
Clinical Trials Unit). To improve training of current and future generations of
statisticians, researchers who are involved in teaching should therefore also
include simulation study methodology in their curricula. The standards of
simulation studies in many statistics related fields (for instance, machine
learning, psychometrics, econometrics, or ecology) are arguably not much
different. One possible avenue for future research is thus to also promote
education and adaptation of simulation study methodology for the special needs
in these fields.

\subsection{Recommendations for editors and reviewers}
Peer review is an important tool for identifying QRPs in research results
submitted to methodological journals. For instance, reviewers may demand
researchers to include competitor methods which are not part of their comparison
yet (or which might have been excluded from the comparison). However, reviewers
can only identify a subset of all QRPs since some types are impossible to spot
if no pre-registered simulation protocol is in place (for example, a reviewer
cannot know whether the evaluation criterion was switched). Even QRPs which can
be detected by peer review may be difficult to spot in practice. It is thus
important that reviewers and editors promote that authors make simulation
protocols and computer code available alongside the manuscript. Moreover, by
providing enough space and encouraging authors to provide detailed descriptions
of their simulation studies, replicability of the simulation studies can be
improved. Finally, reviewers should not be satisfied with manuscripts showing
that a method is uniformly superior; they should also encourage authors to
explore conditions in which their method is expected to be inferior to other
methods or to break down entirely.

\subsection{Recommendations for journals and funding bodies}
Journals and funding bodies can improve on the status quo by either actively
requiring or passively incentivizing more rigorous and neutral simulation study
methodology. Actively, journals can make (pre-registered) simulation protocols
mandatory for all articles featuring a simulation study. A more passive and less
extreme measure would be to indicate with a badge whether an article contains a
pre-registered simulation study, or to introduce article types dedicated to
neutral comparison studies. Such an approach rewards researchers who take the
extra effort. Similar initiatives have led to a large increase in the adoption
of pre-registered study protocols in the field of psychology
\citep{Kidwell2016}. Another measure could be to require standardized reporting
of simulation studies, for example, the ``ADEMP'' reporting structure proposed
by \citet{Morris2019}. Journals may also employ reproducibility checks to ensure
computational reproducibility of the published simulation studies. This is
already done, for example, by the Journal of Open Source Software or the Journal
of Statistical Software. Moreover, journals and funding bodies can promote or
fund research and software to improve simulation study methodology. For
instance, a journal might have special calls for papers on simulation
methodology. Similarly, a funding body could have special grants dedicated to
software development that facilitates sound design, execution and reporting of
simulation studies \citep[as][]{White2010, Gasparini2018, Chalmers2020}.
Finally, journals and funding bodies often exert a strong incentive on
researchers to publish novel and superior methods. This may lead to articles
with non-systematic simulation studies that mainly highlight settings beneficial
to the proposed methods. We believe that the above recommendations can shift the
incentive structure towards more transparent and neutral simulation studies, and
away from the ``one method fits all data sets'' philosophy \citep{Strobl2022}.

%%%%%%%%%%%%%%%%%%%%%%%%%%%%%%%%%%%%%%%%%%%%%%%%%%%%%%%%%%%%%%%%%%%%%%%%%%%%%%%%
\section{Conclusions} \label{sec:discussion}
%%%%%%%%%%%%%%%%%%%%%%%%%%%%%%%%%%%%%%%%%%%%%%%%%%%%%%%%%%%%%%%%%%%%%%%%%%%%%%%%

Simulation studies should be viewed and treated analogously to (empirical)
experiments from other fields of science. Transparent reporting of methodology
and results is essential to contextualize the outcome of such a study. As in
other empirical sciences, QRPs in simulation studies can obfuscate the
usefulness of a novel method and lead to misleading and non-replicable results.

By deliberately using several QRPs we were able to present a method with no
expected benefits and little theoretical justification -- invented solely for
this article -- as an improvement over theoretically and empirically
well-established competitors. While such intentional engagement in these
practices is far from the norm, unintentional QRPs may have the same detrimental
effect. We hope that our illustration will increase awareness about the
fragility of findings from simulation studies and the need for higher standards.

While this article focuses on comparative simulation studies, many of the issues
and recommendations also apply to neutral comparison studies with real data sets
as discussed in \citet{Niessl2021}. Some of the noted problems even exist in
theoretical research; due to the incentive to publish positive results,
researchers often selectively study optimality conditions of methods rather than
conditions under which they fail.

Again, it is imperative to note that researchers rarely engage in QRPs with
malicious intent but because humans tend to interpret ambiguous information
self-servingly, and because they are good at finding reasonable justifications
that match their expectations and desires \citep{Simmons2011}. As in other
domains of science, it is easier to publish positive results in methodological
research, that is, novel and superior methods \citep{Boulesteix2015}. Thus,
methodological researchers will typically desire to show the superiority of a
method rather than to neutrally disclose its strengths and weaknesses.

We provide several recommendations involving various stakeholders in the
research community which we believe may help incentivize researchers to perform
well-designed simulation studies. Most importantly, we think that reviewers,
journals and funders should raise the standards for simulation studies by
promoting pre-registered simulation protocols and rewarding researchers who
invest the extra effort. Although there is evidence for the effectiveness of
protocols in preventing QRPs in other fields, it is unclear whether this effect
translates to simulation studies. Indeed, there are many reasons to believe that
simulation studies will not benefit in a similar way as studies with human or
animal subjects, due to the nature of simulations studies. For instance,
requiring pre-registered protocols cannot prevent researchers engaging in QRPs
until they find their desired results and only then writing and registering a
protocol. In addition, there is currently no tradition of pre-registration in
simulation studies, no best-practices guidance and no dedicated platform to
publish protocols. For example, \citet{Kipruto2022} published the
pre-registration of their simulation protocol as a journal article, whereas the
pre-registration of the protocol from our study was uploaded to GitHub. Both
protocols use the ADEMP reporting structure from \citet{Morris2019}, yet the
field could benefit from reporting guidelines developed by a consortium of
simulation experts similar to the guidelines for health research promoted by the
EQUATOR Network \citep{Altman2008}. Similarly, the field could benefit from a
centralized pre-registration platform tailored to simulation studies (similar to
\url{https://clinicaltrials.gov} for clinical trials). Regardless of the
(unknown) effectiveness of pre-registered simulation protocols, we personally
think that they are an important step toward improving simulation studies since
they promote a minimum degree of transparency and credibility. For this reason,
we think that they are especially important for ``late-stage'' methodological
studies \citep{Heinze2022} where the objective is to neutrally compare different
methods and generate robust evidence.

\section*{Software and data}
The simulation study was conducted in the \textsf{R} language for statistical
computing \citep{pkg:base} using the version 4.1.1. The method \ainet{} is
implemented in the \pkg{ainet} package and available on GitHub
(\url{https://github.com/LucasKook/ainet}). We provide scripts for reproducing
the different simulation studies in the electronic appendix. Due to the
computational overhead, we also provide the resulting data so that the analyses
can be conducted without rerunning the simulations. The data can be downloaded
from Zenodo (\url{https://doi.org/10.5281/zenodo.6364574}). We used \pkg{pROC}
version 1.18.0 to compute the AUC \citep{pkg:proc}. Random forests were fitted
using \pkg{ranger} version 0.13.1 \citep{ranger2017}. For penalized likelihood
methods, we used \pkg{glmnet} version 4.1.2 \citep{Friedman2010,Simon2011}. The
\pkg{SimDesign} package version 2.7.1 was used to set up simulation scenarios
\citep{Chalmers2020}.

\section*{Acknowledgements}
We thank Eva Furrer, Malgorzata Roos and Torsten Hothorn for helpful discussion
and comments on the simulation protocol and drafts of the manuscript. We also
thank the anonymous referees and the associate editor for constructive and
valuable comments that improved the manuscript substantially. Our acknowledgment
of these individuals does not imply their endorsement of this article. The
authors declare that they do not have any conflicts of interest. SP acknowledges
financial support from the Swiss National Science Foundation (Project~\#189295).
The funder had no role in study design, data collection, data analysis, data
interpretation, decision to publish, or preparation of the manuscript.

% Appendix
% ======================================================================
\begin{appendices}

\section{Simulation protocol}
\label{appendix:protocol}

Below, we include an excerpt of the final version of the protocol for the
simulation-based evaluation of \ainet{}. All time-stamped versions of the
protocol are available at \url{https://doi.org/10.5281/zenodo.6364574}.

%%%%%%%%%%%%%%%%%%%%%%%%%%%%%%%%%%%%%%%%%%%%%%%%%%%%%%%%%%%%%%%%%%%%%%%%%%%%%%%%
\subsection{Aims} \label{sec:aims}
%%%%%%%%%%%%%%%%%%%%%%%%%%%%%%%%%%%%%%%%%%%%%%%%%%%%%%%%%%%%%%%%%%%%%%%%%%%%%%%%

The aim of this simulation study is to systematically study the predictive
performance of \ainet{} for a binary prediction task. The simulation conditions
should resemble typical conditions found in the development of prediction models
in biomedical research. In particular we want to evaluate the performance of
\ainet{} conditional on
\begin{itemize}
  \item low- and high-dimensional covariates
  \item (un-)correlated covariates
  \item small and large sample sizes
  \item varying baseline prevalences
\end{itemize}
\ainet{} will be compared to other (penalized) binary regression models
from the literature, namely
\begin{itemize}
  \item Binary logistic regression: the simplest and most popular method for
        binary prediction
  \item Elastic net: a generalization of LASSO and ridge regression, the most
        widely used penalized regression methods
  \item Adaptive elastic net: a generalization of the most popular weighted
  penalized regression method (adaptive LASSO)
  \item Random forest: a popular, more flexible method. This method is related to
        \ainet{}, see Section~\ref{sec:methods}.
\end{itemize}
These cover a wide range of established methods with varying flexibility and
serve as a reasonable benchmark for \ainet. There are many more extensions of
the adaptive elastic net in the literature \citep[see \eg{} the review
by][]{Vidaurre2013}. However, most of these extensions focus on variable
selection and estimation instead of prediction, which is why we restrict our
focus only on the four methods above.

%%%%%%%%%%%%%%%%%%%%%%%%%%%%%%%%%%%%%%%%%%%%%%%%%%%%%%%%%%%%%%%%%%%%%%%%%%%%%%%%
\subsection{Data-generating process} \label{sec:dgp}
%%%%%%%%%%%%%%%%%%%%%%%%%%%%%%%%%%%%%%%%%%%%%%%%%%%%%%%%%%%%%%%%%%%%%%%%%%%%%%%%

In each simulation $b = 1, \dots, B$, we generate a data set consisting of $n$
realizations, \ie $\{(\ry_i, \rx_i)\}_{i=1}^n$. A datum $(\rY, \rX)$ consists of
a binary outcome $\rY \in \{0, 1\}$ and $p$-dimensional covariate vector
$\rX \in \RR^p$. The binary outcomes are generated by
\begin{align*}
  Y \given \rx &\sim \BD\left(\expit\left\{\beta_0 +
	\rx^\top\shiftparm\right\}\right)
\end{align*}
with $\expit(z) = (1 + \exp(-z))^{-1}$ and the covariate vectors are generated by
\begin{align*}
  \rX &\sim \ND_p\left(0, \Sigma\right)
\end{align*}
with covariance matrix $\Sigma$ that may vary across simulation conditions (see
below). The baseline prevalence is $\prev = \expit(\beta_0)$. The coefficient
vector $\shiftparm$ is generated from
\begin{align*}
  \shiftparm \sim \ND_p(0, \Id)
\end{align*}
once per simulation. Finally, the simulation parameters are varied fully
factorially (except for the removal of some unreasonable conditions) as
described below, leading to a total of 128 scenarios.

\subsection*{Sample size}
The sample size used in the development of predictions models varies widely
\citep{Damen2016}. We will use $n \in \{100, 500, 1000, 5000\}$, which span
typical values occurring in practice. Note that previous simulation studies
usually chose sample size based on the implied number of events together with
the number of covariates in the model for easier interpretation
\citep{vanSmeden2018, Riley2018}. We will use this approach in reverse to
determine the dimensionality of the parameters below.

\subsection*{Dimensionality}
Previous simulation studies showed that events per variable ($\EPV$) rather than
the absolute sample size $n$ and dimensionality $p$ influences the predictive
performance of a method. We will therefore define the dimensionality $p$ via EPV
by $$p = \frac{n \cdot \prev}{\EPV}$$ and $2 \leq p \leq 100.$ If the above
formula gives non-integer values, the next larger integer will be used for $p$.
When the formula gives values above 100 or below 2, this simulation condition
will be removed from the design. This is done because prediction models are in
practice only multivariable models ($p \geq 2$), but at the same time the number
of predictors is rarely larger than $p \geq 100$
\citep{Kreuzberger2020,Seker2020, Wynants2020}. The exception are studies
considering complex data, such as images, omics, or text data which are not the
focus here. The values $\EPV \in \{20, 10, 1, 0.5\}$ are chosen to cover
scenarios with small to large number of covariates \citep[see][]{vanSmeden2018}.

\subsection*{Collinearity in $\rX$}
We distinguish between no, low, medium and high collinearity. The diagonal
elements of $\Sigma$ are given by $\Sigma_{ii} = 1$ and the off-diagonal
elements are set to $\Sigma_{ij} = \rho$, $\rho \in \{0, 0.3, 0.6, 0.95\}$.
These values cover the typical (positive) range of correlations.

\subsection*{Baseline prevalence}
Different baseline prevalences $\expit(\beta_0) \in \{0.01, 0.05, 0.1\}$ are
considered, reflecting a reasonable range of prevalences for rare to
common diseases/adverse events.

\subsection*{Test data}
In order to test the out-of-sample predictive performance, we generate a test
data set of $n_{\text{test}} = 10000$ data points in each simulation $b$.

%%%%%%%%%%%%%%%%%%%%%%%%%%%%%%%%%%%%%%%%%%%%%%%%%%%%%%%%%%%%%%%%%%%%%%%%%%%%%%%%
\subsection{Estimands} \label{sec:estimands}
%%%%%%%%%%%%%%%%%%%%%%%%%%%%%%%%%%%%%%%%%%%%%%%%%%%%%%%%%%%%%%%%%%%%%%%%%%%%%%%%
We will estimate different quantities to evaluate overall predictive
performance, calibration, and discrimination, respectively. All methods will be
evaluated on independently generated test data.

\subsubsection{Primary estimand}

\begin{itemize}
  \item \textbf{Brier score.} We compute the Brier score as
  $$\BS = n_{\text{test}}^{-1} \sum_{i=1}^{n_{\text{test}}} (y_{i} - \hat{y}_{i})^2,$$
        where $\hat{y} = \widehat\Prob(Y = 1 \given \rx)$. Lower values indicate
        better predictive performance in terms of calibration and sharpness. A
        prediction is well-calibrated if the observed proportion of events is
        close to the predicted probabilities. Sharpness refers to how
        concentrated a predictive distribution is (\eg{} how wide/narrow a
        prediction interval is), and the predictive goal is to maximize
        sharpness subject to calibration \citep{Gneiting2008}. The Brier score
        is a proper scoring rule, meaning that it is minimized if a predicted
        distribution is equal to the data-generating distribution
        \citep{Gneiting2007}. Proper scoring rules thus encourage honest
        predictions. The Brier score is therefore a principled choice for our
        primary estimand.
\end{itemize}

\subsubsection{Secondary estimands}
\begin{itemize}
  \item \textbf{Scaled Brier score.} The scaled Brier score (also known as Brier
        skill score) is computed as
  $$\BS^{*} = 1 - \BS/\BS_{0}$$
        with $\BS_{0} = \bar{y}(1 - \bar{y})$ and $\bar{y}$ the observed
        prevalence in the data set. The scaled Brier score takes into account
        that the prevalence varies across simulation conditions. Hence, the
        scaled Brier score can be compared between conditions \citep{Schmid2005,
        steyerberg2019clinical}.

  \item \textbf{Log-score.} We compute the log-score on independently generated test data,
  $$\LS = - n_{\text{test}}^{-1} \sum_{i=1}^{n_{\text{test}}} \left\{ y_{i} \log(\hat{y}_{i})
        + (1 - y_{i}) \log (1 - \hat{y}_{i})\right\},$$ will be used as a
        secondary measure of overall predictive performance. Lower values
        indicate better predictive performance in terms of calibration and
        sharpness. The log-score is a strictly proper scoring rule, however, it
        is more sensitive to extreme predicted probabilities compared to the
        Brier score \citep{Gneiting2007}.

  \item \textbf{AUC.} The AUC is the area under the
        receiver-operating-characteristic (ROC) curve
        \citep{steyerberg2019clinical}. It will be used as a measure of
        discrimination and values closer to one indicate better discriminative
        ability. Discrimination describes the ability of a prediction model to
        discriminate between cases and non-cases. Other discrimination measures,
        such as accuracy, sensitivity, specificity, etc., are not considered
        because we want to evaluate predictive performance in terms of
        probabilistic predictions instead of point predictions/classification.

  \item \textbf{Calibration slope $\hat b$.} The calibration slope $\hat b$ is
        obtained by regressing the test data outcomes $y_{\text{test}}$ on the
        models' predicted logits $\logit({\hat{y}})$, \ie
  $$\logit\Ex[Y \given \hat\ry] = a + b\logit(\hat\ry).$$
        This measure will be used to assess calibration and deviations of
        $\hat b$ from one indicate miscalibration
        \citep{steyerberg2019clinical}.

  \item \textbf{Calibration in the large $\hat a$.} We inspect calibration in
        the large $\hat a$ on independently generated test data, from the model
  $$\logit\Ex[Y \given \hat\ry] = a + \logit(\hat{y}).$$
        This measure will also be used to assess calibration and deviations of
        $\hat a$ from zero indicate miscalibration
        \citep{steyerberg2019clinical}.
\end{itemize}

To facilitate comparison between simulation conditions, all estimands will also
be corrected by the oracle version of the estimand, \eg{} the Brier score will
be computed from the ground truth parameters and the simulated data $\rx$,
subsequently the oracle Brier score will be subtracted from the estimated Brier
score.

%%%%%%%%%%%%%%%%%%%%%%%%%%%%%%%%%%%%%%%%%%%%%%%%%%%%%%%%%%%%%%%%%%%%%%%%%%%%%%%%
\subsection{Methods} \label{sec:methods}
%%%%%%%%%%%%%%%%%%%%%%%%%%%%%%%%%%%%%%%%%%%%%%%%%%%%%%%%%%%%%%%%%%%%%%%%%%%%%%%%

\subsubsection{\ainet}
We now present the mock-method and give a superficial motivation why it could
lead to improved predictive performance: Choosing the vector of penalization
weights in the adaptive LASSO becomes difficult in high-dimensional settings.
For instance, using absolute LASSO estimates as penalization weights omits the
importance of several predictors by not selecting them, especially in the case
of highly correlated predictors \citep{Algamal2015}. The adaptive importance
elastic net (\ainet{}) circumvents this problem by employing a random forest to
estimate the penalization weights via an \emph{a priori} chosen variable
importance measure. In this way, the importance of all variables enter the
penalization weights simultaneously.

The penalized log-likelihood for \ainet{} for a single observation $(\ry, \rx)$
is defined as
$$\ell_{\text{AINET}}(\beta_0, \shiftparm; \ry, \rx, \alpha, \lambda, \wvec) =
  \ell(\beta_0, \shiftparm; \ry, \rx) + \aipen$$
where
$$\ell(\beta_0, \shiftparm; \ry, \rx) =
\ry \log\left(\expit\left\{\beta_0 + \linpred\right\}\right) + (1 - \ry) \log\left(1 - \expit\left\{\beta_0 + \linpred\right\}\right)$$
denotes the log-likelihood of a binomial GLM and $\wvec$ is derived from a
random forest variable importance measure $\widetilde\IMP$ as
$$w_j = 1 - \left(\frac{\IMP_j}{\sum_{k=1}^p \IMP_k}\right)^\gamma,$$
where we transform $\IMP$ to be non-negative via
% $$\IMP = \widetilde\IMP - \min_j\{\widetilde{\IMP}_j\},$$
% or
$$\IMP_j = \max\{0, \widetilde\IMP_j\}$$
and $\gamma$ is a hyperparameter for the influence of the weights similar to
$\gamma$ hyperparameter of the adaptive elastic net. \ainet{} is fitted by
maximizing its penalized log-likelihood assuming i.i.d. observations
$\{(\ry_i, \rx_i)\}_{i=1}^n$, \ie
$$\arg\max_{\beta_0, \shiftparm} \sum_{i = 1}^n \ell_{\text{AINET}}(\beta_0,
\shiftparm; \ry_i, \rx_i, \alpha, \lambda, \wvec).$$

Per default, we choose mean decrease in the Gini coefficient for
$\widetilde\IMP$. Hyperparameters of the random forest are not tuned, but kept
at their default values (\eg{} \code{mtry}, \code{ntree}). The hyperparameter
$\gamma = 1$ will stay constant for all simulations.

\ainet{} is supposed to seem like a reasonable method at first glance. However,
\ainet{} cannot be expected to share desirable theoretical properties with the
usual adaptive LASSO, such as oracle estimation \citep{Zou2006}. This is because
the penalization weights $\wvec$ do not meet the required consistency
assumption. Also in terms of prediction performance, \ainet{} is not expected to
outperform methods of comparable complexity.

\subsubsection{Benchmark methods}

\begin{itemize}
  \item \textbf{Binary logistic regression} \citep{mccullagh2019generalized}
        with and without ridge penalty for high- and low-dimensional settings,
        respectively. In case a ridge penalty is needed, it is tuned via 5-fold
        cross-validation by following the ``one standard error'' rule as
        implemented in \pkg{glmnet} \citep{Friedman2010}.
  \item \textbf{Elastic net} \citep{Zou2005}, for which the penalized
        log-likelihood is given by
    $$\ell_{\text{EN}}(\beta_0, \shiftparm; \ry, \rx, \alpha, \lambda) =
        \ell(\beta_0, \shiftparm; \ry, \rx) + \llpen.$$ Here, $\alpha$ and
        $\lambda$ are tuned via 5-fold cross-validation by following the ``one
        standard error'' rule.
  \item \textbf{Adaptive elastic net} \citep{Zou2006}, with penalized loss
        function
    $$\ell_{\text{adaptive}}(\beta_0, \shiftparm; \ry, \rx, \alpha, \lambda, \wvec)
        = \ell(\beta_0, \shiftparm; \ry, \rx) + \aipen.$$ Here, the penalty
        weights $\wvec$ are inverse coefficient estimates from a binary logistic
        regression
    $$\hat{w}_j = \lvert\hat\eparm_j\rvert^{-\gamma},$$
        where $\lambda$ and $\alpha$ are tuned via 5-fold cross-validation by
        following the ``one standard error'' rule. The hyperparameter
        $\gamma = 1$ will stay constant for all simulations. In case $p > n$, we
        estimate the penalty weights using a ridge penalty, tuned via an
        additional nested 5-fold cross-validation by following the ``one
        standard error'' rule.
  \item \textbf{Random forests} \citep{Breiman2001} for binary outcomes without
        hyperparameter tuning. The default parameters of \pkg{ranger} will be
        used \citep{ranger2017}. % , \ie
    % \code{ntree = 500}, \code{mtry = floor(sqrt(p))}, \code{min.node.size = },
    % \code{max.depth = }, \code{sample.fraction = }
\end{itemize}

%%%%%%%%%%%%%%%%%%%%%%%%%%%%%%%%%%%%%%%%%%%%%%%%%%%%%%%%%%%%%%%%%%%%%%%%%%%%%%%%
\subsection{Performance measures} \label{sec:performance}
%%%%%%%%%%%%%%%%%%%%%%%%%%%%%%%%%%%%%%%%%%%%%%%%%%%%%%%%%%%%%%%%%%%%%%%%%%%%%%%%

The distribution of all estimands from Section~\ref{sec:estimands} will be
assessed visually with box- and violin-plots that are stratified by method and
simulation conditions. We will also compute mean, median, standard deviation,
interquartile range, and 95\% confidence intervals for each of the estimands.
Moreover, instead of ``eye-balling'' differences in predictive performance
across methods and conditions, we will formally assess them by regressing the
estimands on the method and simulation conditions \citep[\cf][]{Skrondal2000}.
To do so, we will use a fully interacted model with the interaction between the
methods and the 128 simulations conditions, \ie in R notation: \texttt{estimand
  $\sim$ 0 + method:scenario}. We will rank pairwise comparison between two
methods within a single condition by their $p$-values, to more easily identify
conditions where methods show differences in predictive performance. The choice
of a significance level at which a method is deemed superior will be determined
based on preliminary simulations. We set this level to 5\%, where $p$-values
will be adjusted using the single-step method \citep{pkg:multcomp} within a
single simulation condition for comparisons between \ainet{} and any other
method.

\subsection{Determining the number of simulations}

We determine the number of simulation $B$ such that the Monte Carlo standard
error of the primary estimand, the mean Brier score $\BS/B$, is sufficiently
small. The variance of $\BS/B$ is given by
\begin{align*}
  \Var\left(\BS/B\right)
  &= \frac{\Var\left\{(y -
    \hat{y})^{2}\right\}}{B \cdot n_{\text{test}}}
  % &= B^{-1}n_{\text{test}}^{-1} \left\{\Ex[(y_{ib} - \hat{y}_{ib})^{4}] -
  %   \Ex[(y_{ib} - \hat{y}_{ib})^{2}]^{2}\right\}
\end{align*}
and $\Var\left\{(y_{ib} - \hat{y}_{ib})^{2}\right\}$ could be decomposed further
\citep{Bradley2008}. However, the resulting expression is difficult to evaluate
for our data-generating process as it depends on several of the simulation
parameters. We therefore follow a similar approach as in \citet{Morris2019} and
estimate $\widehat{\Var}\left\{(y_{ib} - \hat{y}_{ib})^{2}\right\} < V$ from an
initial small simulation run with 100 simulations per condtion to get an upper
bound $V$ for worst-case variance across all simulation conditions. Therefore,
the number of simulations is then given by
$$B = \frac{V}{n_{\text{test}} \Var\left(\BS\right)}.$$
Since $\BS \in [0, 1]$ we decide that we require the Monte Carlo standard error
of $\BS$ to be lower than four significant digits, $0.0001$.

The initial simulation run led to an estimated worst case variance of
$\widehat{V} = 0.2$. Therefore, we compute that
$$B = 0.2/(10000 \times 0.0001^{2}) = 2000$$
replications are required to obtain Brier score estimates with the
desired precision.

%%%%%%%%%%%%%%%%%%%%%%%%%%%%%%%%%%%%%%%%%%%%%%%%%%%%%%%%%%%%%%%%%%%%%%%%%%%%%%%%
\subsection{Handling exceptions} \label{sec:exceptions}
%%%%%%%%%%%%%%%%%%%%%%%%%%%%%%%%%%%%%%%%%%%%%%%%%%%%%%%%%%%%%%%%%%%%%%%%%%%%%%%%
It is inevitable that convergence issues and other problems will arise in the
simulation study. We will handle them as follows:
\begin{itemize}
  \item If a method fails to converge, the simulation will be excluded from the
        analysis. The failing simulations will not be replaced with new
        simulations that successfully converge as convergence may be impossible
        for some scenarios.
  \item We will report the proportion of simulations with convergence issues for
        each method and discuss the potential reasons for their emergence.
  \item In case of severe convergence issues or other problems (more than 10\%
        of the simulations failing within a setting), we may adjust the
        simulation parameters post hoc. This will be indicated in the discussion
        of the results.
  \item Convergence may be possible for certain tuning parameters of a method
        (\eg{} cross-validation of LASSO may fail for some values $\lambda$
        while it could work for others). In this case we will choose a parameter
        value where the method still converges, as one would usually do with a
        real data set.
\end{itemize}

\end{appendices}

% Bibliography
% ======================================================================
\bibliographystyle{apalikedoiurl}
\bibliography{bibliography}

\begin{thebibliography}{}

\bibitem[Algamal and Lee, 2015]{Algamal2015}
Algamal, Z.~Y. and Lee, M.~H. (2015).
\newblock {Penalized logistic regression with the adaptive LASSO for gene
  selection in high-dimensional cancer classification}.
\newblock {\em Expert Systems with Applications}, 42(23):9326--9332.
\newblock \doi{10.1016/J.ESWA.2015.08.016}.

\bibitem[Altman et~al., 2017]{Altman2017}
Altman, D.~G., Moher, D., and Schulz, K.~F. (2017).
\newblock Harms of outcome switching in reports of randomised trials: {CONSORT}
  perspective.
\newblock {\em {BMJ}}, 356:j396.
\newblock \doi{10.1136/bmj.j396}.

\bibitem[Altman et~al., 2008]{Altman2008}
Altman, D.~G., Simera, I., Hoey, J., Moher, D., and Schulz, K. (2008).
\newblock {EQUATOR}: Reporting guidelines for health research.
\newblock {\em The Lancet}, 371(9619):1149--1150.
\newblock \doi{10.1016/s0140-6736(08)60505-x}.

\bibitem[Angelis et~al., 2004]{DeAngelis2004}
Angelis, C.~D., Drazen, J.~M., Frizelle, F.~A., Haug, C., Hoey, J., Horton, R.,
  Kotzin, S., Laine, C., Marusic, A., Overbeke, A. J.~P., Schroeder, T.~V.,
  Sox, H.~C., and Weyden, M. B. V.~D. (2004).
\newblock Clinical trial registration: A statement from the international
  committee of medical journal editors.
\newblock {\em New England Journal of Medicine}, 351(12):1250--1251.
\newblock \doi{10.1056/nejme048225}.

\bibitem[Boulesteix et~al., 2017]{Boulesteix2017b}
Boulesteix, A.-L., Binder, H., Abrahamowicz, M., and Sauerbrei, W. (2017).
\newblock On the necessity and design of studies comparing statistical methods.
\newblock {\em Biometrical Journal}, 60(1):216--218.
\newblock \doi{10.1002/bimj.201700129}.

\bibitem[Boulesteix et~al., 2020]{Boulesteix2020B}
Boulesteix, A.-L., Groenwold, R.~H., Abrahamowicz, M., Binder, H., Briel, M.,
  Hornung, R., Morris, T.~P., Rahnenf\"{u}hrer, J., and Sauerbrei, W. (2020).
\newblock Introduction to statistical simulations in health research.
\newblock {\em {BMJ} Open}, 10(12):e039921.
\newblock \doi{10.1136/bmjopen-2020-039921}.

\bibitem[Boulesteix et~al., 2013]{Boulesteix2013}
Boulesteix, A.-L., Lauer, S., and Eugster, M. J.~A. (2013).
\newblock A plea for neutral comparison studies in computational sciences.
\newblock {\em {PLOS} {ONE}}, 8(4):e61562.
\newblock \doi{10.1371/journal.pone.0061562}.

\bibitem[Boulesteix et~al., 2015]{Boulesteix2015}
Boulesteix, A.-L., Stierle, V., and Hapfelmeier, A. (2015).
\newblock Publication bias in methodological computational research.
\newblock {\em Cancer Informatics}, 14s5:CIN.S30747.
\newblock \doi{10.4137/cin.s30747}.

\bibitem[Bradley et~al., 2008]{Bradley2008}
Bradley, A.~A., Schwartz, S.~S., and Hashino, T. (2008).
\newblock Sampling uncertainty and confidence intervals for the {Brier} score
  and {Brier} skill score.
\newblock {\em Weather and Forecasting}, 23(5):992--1006.
\newblock \doi{10.1175/2007waf2007049.1}.

\bibitem[Breiman, 2001]{Breiman2001}
Breiman, L. (2001).
\newblock {Random Forests}.
\newblock {\em Machine Learning}, 45(1):5--32.
\newblock \doi{10.1023/a:1010933404324}.

\bibitem[Burton et~al., 2006]{Burton2006}
Burton, A., Altman, D.~G., Royston, P., and Holder, R.~L. (2006).
\newblock The design of simulation studies in medical statistics.
\newblock {\em Statistics in Medicine}, 25(24):4279--4292.
\newblock \doi{10.1002/sim.2673}.

\bibitem[Chalmers and Adkins, 2020]{Chalmers2020}
Chalmers, R.~P. and Adkins, M.~C. (2020).
\newblock Writing effective and reliable {Monte Carlo} simulations with the
  {SimDesign} package.
\newblock {\em The Quantitative Methods for Psychology}, 16(4):248--280.
\newblock \doi{10.20982/tqmp.16.4.p248}.

\bibitem[Chipman and Bingham, 2022]{Chipman2022}
Chipman, H. and Bingham, D. (2022).
\newblock Let's practice what we preach: Planning and interpreting simulation
  studies with design and analysis of experiments.
\newblock {\em Canadian Journal of Statistics}, 50(4):1228--1249.
\newblock \doi{10.1002/cjs.11719}.

\bibitem[Damen et~al., 2016]{Damen2016}
Damen, J. A. A.~G., Hooft, L., Schuit, E., Debray, T. P.~A., Collins, G.~S.,
  Tzoulaki, I., Lassale, C.~M., Siontis, G. C.~M., Chiocchia, V., Roberts, C.,
  Schl\"{u}ssel, M.~M., Gerry, S., Black, J.~A., Heus, P., van~der Schouw,
  Y.~T., Peelen, L.~M., and Moons, K. G.~M. (2016).
\newblock Prediction models for cardiovascular disease risk in the general
  population: Systematic review.
\newblock {\em {BMJ}}, 353:i2416.
\newblock \doi{10.1136/bmj.i2416}.

\bibitem[Dutilh et~al., 2021]{Dutilh2019}
Dutilh, G., Sarafoglou, A., and Wagenmakers, E.-J. (2021).
\newblock Flexible yet fair: Blinding analyses in experimental psychology.
\newblock {\em Synthese}, 198(Suppl 23):5745--5772.
\newblock \doi{10.1007/s11229-019-02456-7}.

\bibitem[Elofsson et~al., 2019]{Elofsson2019}
Elofsson, A., Hess, B., Lindahl, E., Onufriev, A., van~der Spoel, D., and
  Wallqvist, A. (2019).
\newblock Ten simple rules on how to create open access and reproducible
  molecular simulations of biological systems.
\newblock {\em {PLOS} Computational Biology}, 15(1):e1006649.
\newblock \doi{10.1371/journal.pcbi.1006649}.

\bibitem[Friedman et~al., 2010]{Friedman2010}
Friedman, J., Hastie, T., and Tibshirani, R. (2010).
\newblock {Regularization Paths for Generalized Linear Models via Coordinate
  Descent}.
\newblock {\em Journal of Statistical Software}, 33(1):1--22.
\newblock \doi{10.18637/jss.v033.i01}.

\bibitem[Gasparini, 2018]{Gasparini2018}
Gasparini, A. (2018).
\newblock rsimsum: Summarise results from {Monte} {Carlo} simulation studies.
\newblock {\em Journal of Open Source Software}, 3(26):739.
\newblock \doi{10.21105/joss.00739}.

\bibitem[Gasparini et~al., 2021]{Gasparini2021}
Gasparini, A., Morris, T.~P., and Crowther, M.~J. (2021).
\newblock {INTEREST}: {INteractive} tool for exploring {REsults} from
  simulation {sTudies}.
\newblock {\em Journal of Data Science, Statistics, and Visualisation}, 1(4).
\newblock \doi{10.52933/jdssv.v1i4.9}.

\bibitem[Gelman and Tuerlinckx, 2000]{Gelman2000}
Gelman, A. and Tuerlinckx, F. (2000).
\newblock Type {S} error rates for classical and {Bayesian} single and multiple
  comparison procedures.
\newblock {\em Computational Statistics}, 15(3):373--390.
\newblock \doi{10.1007/s001800000040}.

\bibitem[Gneiting, 2008]{Gneiting2008}
Gneiting, T. (2008).
\newblock Editorial: Probabilistic forecasting.
\newblock {\em Journal of the Royal Statistical Society: Series A (Statistics
  in Society)}, 171(2):319--321.
\newblock \doi{10.1111/j.1467-985x.2007.00522.x}.

\bibitem[Gneiting and Raftery, 2007]{Gneiting2007}
Gneiting, T. and Raftery, A.~E. (2007).
\newblock Strictly proper scoring rules, prediction, and estimation.
\newblock {\em Journal of the American Statistical Association},
  102(477):359--378.
\newblock \doi{10.1198/016214506000001437}.

\bibitem[Heinze et~al., 2023]{Heinze2022}
Heinze, G., Boulesteix, A.-L., Kammer, M., Morris, T.~P., and and, I. R.~W.
  (2023).
\newblock Phases of methodological research in biostatistics--building the
  evidence base for new methods.
\newblock {\em Biometrical Journal}, page 2200222.
\newblock \doi{10.1002/bimj.202200222}.

\bibitem[Hoaglin and Andrews, 1975]{Hoaglin1975}
Hoaglin, D.~C. and Andrews, D.~F. (1975).
\newblock The reporting of computation-based results in statistics.
\newblock {\em The American Statistician}, 29(3):122--126.
\newblock \doi{10.1080/00031305.1975.10477393}.

\bibitem[Hoffmann et~al., 2021]{Hoffmann2021}
Hoffmann, S., Sch\"{o}nbrodt, F., Elsas, R., Wilson, R., Strasser, U., and
  Boulesteix, A.-L. (2021).
\newblock The multiplicity of analysis strategies jeopardizes replicability:
  Lessons learned across disciplines.
\newblock {\em Royal Society Open Science}, 8(4).
\newblock \doi{10.1098/rsos.201925}.

\bibitem[Holford et~al., 2000]{Holford2000}
Holford, N. H.~G., Kimko, H.~C., Monteleone, J. P.~R., and Peck, C.~C. (2000).
\newblock Simulation of clinical trials.
\newblock {\em Annual Review of Pharmacology and Toxicology}, 40(1):209--234.
\newblock \doi{10.1146/annurev.pharmtox.40.1.209}.

\bibitem[Hothorn et~al., 2008]{pkg:multcomp}
Hothorn, T., Bretz, F., and Westfall, P. (2008).
\newblock Simultaneous inference in general parametric models.
\newblock {\em Biometrical Journal}, 50(3):346--363.
\newblock \doi{10.1002/bimj.200810425}.

\bibitem[Jelizarow et~al., 2010]{Jelizarow2010}
Jelizarow, M., Guillemot, V., Tenenhaus, A., Strimmer, K., and Boulesteix,
  A.-L. (2010).
\newblock Over-optimism in bioinformatics: An illustration.
\newblock {\em Bioinformatics}, 26(16):1990--1998.
\newblock \doi{10.1093/bioinformatics/btq323}.

\bibitem[Kidwell et~al., 2016]{Kidwell2016}
Kidwell, M.~C., Lazarevi{\'{c}}, L.~B., Baranski, E., Hardwicke, T.~E.,
  Piechowski, S., Falkenberg, L.-S., Kennett, C., Slowik, A., Sonnleitner, C.,
  Hess-Holden, C., Errington, T.~M., Fiedler, S., and Nosek, B.~A. (2016).
\newblock Badges to acknowledge open practices: A simple, low-cost, effective
  method for increasing transparency.
\newblock {\em {PLOS} Biology}, 14(5):e1002456.
\newblock \doi{10.1371/journal.pbio.1002456}.

\bibitem[Kipruto and Sauerbrei, 2022]{Kipruto2022}
Kipruto, E. and Sauerbrei, W. (2022).
\newblock Comparison of variable selection procedures and investigation of the
  role of shrinkage in linear regression-protocol of a simulation study in
  low-dimensional data.
\newblock {\em {PLOS} {ONE}}, 17(10):e0271240.
\newblock \doi{10.1371/journal.pone.0271240}.

\bibitem[Klein and Roodman, 2005]{Klein2005}
Klein, J.~R. and Roodman, A. (2005).
\newblock Blind analysis in nuclear and particle physics.
\newblock {\em Annual Review of Nuclear and Particle Science}, 55(1):141--163.
\newblock \doi{10.1146/annurev.nucl.55.090704.151521}.

\bibitem[Kreutz et~al., 2020]{Kreutz2020}
Kreutz, C., Can, N.~S., Bruening, R.~S., Meyberg, R., M{\'{e}}rai, Z.,
  Fernandez-Pozo, N., and Rensing, S.~A. (2020).
\newblock A blind and independent benchmark study for detecting differentially
  methylated regions in plants.
\newblock {\em Bioinformatics}, 36(11):3314--3321.
\newblock \doi{10.1093/bioinformatics/btaa191}.

\bibitem[Kreuzberger et~al., 2020]{Kreuzberger2020}
Kreuzberger, N., Damen, J., Trivella, M., Estcourt, L.~J., Aldin, A., Umlauff,
  L., Vazquez-Montes, M., Wolff, R., Moons, K., Monsef, I., Foroutan, F.,
  Kreuzer, K., and Skoetz, N. (2020).
\newblock Prognostic models for newly-diagnosed chronic lymphocytic leukaemia
  in adults: A systematic review and meta-analysis.
\newblock {\em Cochrane Database of Systematic Reviews}, 7:CD012022.
\newblock \doi{10.1002/14651858.CD012022.pub2}.

\bibitem[Lawlor, 2007]{Lawlor2007}
Lawlor, D.~A. (2007).
\newblock Quality in epidemiological research: Should we be submitting papers
  before we have the results and submitting more hypothesis-generating
  research?
\newblock {\em International Journal of Epidemiology}, 36(5):940--943.
\newblock \doi{10.1093/ije/dym168}.

\bibitem[Loder et~al., 2010]{Loder2010}
Loder, E., Groves, T., and MacAuley, D. (2010).
\newblock Registration of observational studies.
\newblock {\em {BMJ}}, 340:c950.
\newblock \doi{10.1136/bmj.c950}.

\bibitem[McCullagh and Nelder, 2019]{mccullagh2019generalized}
McCullagh, P. and Nelder, J.~A. (2019).
\newblock {\em {Generalized Linear Models}}.
\newblock Routledge.

\bibitem[Monks et~al., 2018]{Monks2018}
Monks, T., Currie, C. S.~M., Onggo, B.~S., Robinson, S., Kunc, M., and Taylor,
  S. J.~E. (2018).
\newblock Strengthening the reporting of empirical simulation studies:
  Introducing the {STRESS} guidelines.
\newblock {\em Journal of Simulation}, 13(1):55--67.
\newblock \doi{10.1080/17477778.2018.1442155}.

\bibitem[Morris et~al., 2019]{Morris2019}
Morris, T.~P., White, I.~R., and Crowther, M.~J. (2019).
\newblock Using simulation studies to evaluate statistical methods.
\newblock {\em Statistics in Medicine}, 38(11):2074--2102.
\newblock \doi{10.1002/sim.8086}.

\bibitem[Nie{\ss}l et~al., 2022]{Niessl2021}
Nie{\ss}l, C., Herrmann, M., Wiedemann, C., Casalicchio, G., and Boulesteix,
  A.-L. (2022).
\newblock Over-optimism in benchmark studies and the multiplicity of design and
  analysis options when interpreting their results.
\newblock {\em {WIREs} Data Mining and Knowledge Discovery}, 12(2):e1441.
\newblock \doi{10.1002/widm.1441}.

\bibitem[Nosek et~al., 2018]{Nosek2018}
Nosek, B.~A., Ebersole, C.~R., DeHaven, A.~C., and Mellor, D.~T. (2018).
\newblock The preregistration revolution.
\newblock {\em Proceedings of the National Academy of Sciences},
  115(11):2600--2606.
\newblock \doi{10.1073/pnas.1708274114}.

\bibitem[O{\textquotesingle}Kelly et~al., 2016]{OKelly2016}
O{\textquotesingle}Kelly, M., Anisimov, V., Campbell, C., and Hamilton, S.
  (2016).
\newblock Proposed best practice for projects that involve modelling and
  simulation.
\newblock {\em Pharmaceutical Statistics}, 16(2):107--113.
\newblock \doi{10.1002/pst.1789}.

\bibitem[{R Core Team}, 2020]{pkg:base}
{R Core Team} (2020).
\newblock {\em R: A Language and Environment for Statistical Computing}.
\newblock R Foundation for Statistical Computing, Vienna, Austria.
\newblock URL \url{https://www.R-project.org/}.

\bibitem[Riley et~al., 2018]{Riley2018}
Riley, R.~D., Snell, K.~I., Ensor, J., Burke, D.~L., Jr, F. E.~H., Moons,
  K.~G., and Collins, G.~S. (2018).
\newblock Minimum sample size for developing a multivariable prediction model:
  {PART} {II} -- binary and time-to-event outcomes.
\newblock {\em Statistics in Medicine}, 38(7):1276--1296.
\newblock \doi{10.1002/sim.7992}.

\bibitem[Robertson et~al., 2023]{Robertson2022}
Robertson, D.~S., Choodari-Oskooei, B., Dimairo, M., Flight, L., Pallmann, P.,
  and Jaki, T. (2023).
\newblock Point estimation for adaptive trial designs {I}: A methodological
  review.
\newblock {\em Statistics in Medicine}, 42(2):122--145.
\newblock \doi{10.1002/sim.9605}.

\bibitem[Robin et~al., 2011]{pkg:proc}
Robin, X., Turck, N., Hainard, A., Tiberti, N., Lisacek, F., Sanchez, J.-C.,
  and Müller, M. (2011).
\newblock {pROC: An open-source package for R and S+ to analyze and compare ROC
  curves}.
\newblock {\em BMC Bioinformatics}, 12:77.
\newblock \doi{10.1186/1471-2105-12-77}.

\bibitem[Schmid and Griffith, 2005]{Schmid2005}
Schmid, C.~H. and Griffith, J.~L. (2005).
\newblock Multivariate classification rules: Calibration and discrimination.
\newblock In Armitage, P. and Colton, T., editors, {\em Encyclopedia of
  Biostatistics}, volume~5, pages 3491--3497. Wiley, 2nd edition.

\bibitem[Schwab and Held, 2021]{Schwab2021}
Schwab, S. and Held, L. (2021).
\newblock Statistical programming: Small mistakes, big impacts.
\newblock {\em Significance}, 18(3):6--7.
\newblock \doi{10.1111/1740-9713.01522}.

\bibitem[Seker et~al., 2020]{Seker2020}
Seker, B.~O., Reeve, K., Havla, J., Burns, J., Gosteli, M., Lutterotti, A.,
  Schippling, S., Mansmann, U., and Held, U. (2020).
\newblock Prognostic models for predicting clinical disease progression,
  worsening and activity in people with multiple sclerosis.
\newblock {\em Cochrane Database of Systematic Reviews}, (5).
\newblock \doi{10.1002/14651858.CD013606}.

\bibitem[Simmons et~al., 2011]{Simmons2011}
Simmons, J.~P., Nelson, L.~D., and Simonsohn, U. (2011).
\newblock False-positive psychology: Undisclosed flexibility in data collection
  and analysis allows presenting anything as significant.
\newblock {\em Psychological Science}, 22(11):1359--1366.
\newblock \doi{10.1177/0956797611417632}.

\bibitem[Simon et~al., 2011]{Simon2011}
Simon, N., Friedman, J., Hastie, T., and Tibshirani, R. (2011).
\newblock {Regularization Paths for Cox's Proportional Hazards Model via
  Coordinate Descent}.
\newblock {\em Journal of Statistical Software}, 39(5):1--13.
\newblock \doi{10.18637/jss.v039.i05}.

\bibitem[Skrondal, 2000]{Skrondal2000}
Skrondal, A. (2000).
\newblock Design and analysis of {Monte} {Carlo} experiments: Attacking the
  conventional wisdom.
\newblock {\em Multivariate Behavioral Research}, 35(2):137--167.
\newblock \doi{10.1207/s15327906mbr3502_1}.

\bibitem[Smith and Marshall, 2010]{Smith2010}
Smith, M.~K. and Marshall, A. (2010).
\newblock Importance of protocols for simulation studies in clinical drug
  development.
\newblock {\em Statistical Methods in Medical Research}, 20(6):613--622.
\newblock \doi{10.1177/0962280210378949}.

\bibitem[Steyerberg et~al., 2019]{steyerberg2019clinical}
Steyerberg, E.~W. et~al. (2019).
\newblock {\em Clinical Prediction Models}.
\newblock Springer.

\bibitem[Strobl and Leisch, 2022]{Strobl2022}
Strobl, C. and Leisch, F. (2022).
\newblock Against the {\textquotedblleft}one method fits all data
  sets{\textquotedblright} philosophy for comparison studies in methodological
  research.
\newblock {\em Biometrical Journal}, 00:1--8.
\newblock \doi{10.1002/bimj.202200104}.

\bibitem[Tukey, 1980]{Tukey1980}
Tukey, J.~W. (1980).
\newblock We need both exploratory and confirmatory.
\newblock {\em The American Statistician}, 34(1):23--25.
\newblock \doi{10.1080/00031305.1980.10482706}.

\bibitem[Ullmann et~al., 2022]{Ullmann2022}
Ullmann, T., Beer, A., H\"{u}nem\"{o}rder, M., Seidl, T., and Boulesteix, A.-L.
  (2022).
\newblock Over-optimistic evaluation and reporting of novel cluster algorithms:
  An illustrative study.
\newblock {\em Advances in Data Analysis and Classification}.
\newblock \doi{10.1007/s11634-022-00496-5}.

\bibitem[Van~der Bles et~al., 2019]{van2019communicating}
Van~der Bles, A.~M., Van Der~Linden, S., Freeman, A.~L., Mitchell, J., Galvao,
  A.~B., Zaval, L., and Spiegelhalter, D.~J. (2019).
\newblock Communicating uncertainty about facts, numbers and science.
\newblock {\em Royal Society Open Science}, 6(5):181870.
\newblock \doi{10.1098/rsos.181870}.

\bibitem[van Smeden et~al., 2016]{vanSmeden2016}
van Smeden, M., de~Groot, J. A.~H., Moons, K. G.~M., Collins, G.~S., Altman,
  D.~G., Eijkemans, M. J.~C., and Reitsma, J.~B. (2016).
\newblock No rationale for 1 variable per 10 events criterion for binary
  logistic regression analysis.
\newblock {\em {BMC} Medical Research Methodology}, 16:163.
\newblock \doi{10.1186/s12874-016-0267-3}.

\bibitem[van Smeden et~al., 2018]{vanSmeden2018}
van Smeden, M., Moons, K.~G., de~Groot, J.~A., Collins, G.~S., Altman, D.~G.,
  Eijkemans, M.~J., and Reitsma, J.~B. (2018).
\newblock Sample size for binary logistic prediction models: Beyond events per
  variable criteria.
\newblock {\em Statistical Methods in Medical Research}, 28(8):2455--2474.
\newblock \doi{10.1177/0962280218784726}.

\bibitem[van Zwet and Cator, 2021]{Vanzwet2021}
van Zwet, E.~W. and Cator, E.~A. (2021).
\newblock The significance filter, the winner's curse and the need to shrink.
\newblock {\em Statistica Neerlandica}, 75(4):437--452.
\newblock \doi{10.1111/stan.12241}.

\bibitem[Vidaurre et~al., 2013]{Vidaurre2013}
Vidaurre, D., Bielza, C., and Larra{\~{n}}aga, P. (2013).
\newblock A survey of {$L_1$} regression.
\newblock {\em International Statistical Review}, 81(3):361--387.
\newblock \doi{10.1111/insr.12023}.

\bibitem[White, 2010]{White2010}
White, I.~R. (2010).
\newblock Simsum: Analyses of simulation studies including monte carlo error.
\newblock {\em The Stata Journal: Promoting communications on statistics and
  Stata}, 10(3):369--385.
\newblock \doi{10.1177/1536867x1001000305}.

\bibitem[Wicherts et~al., 2016]{Wicherts2016}
Wicherts, J.~M., Veldkamp, C. L.~S., Augusteijn, H. E.~M., Bakker, M., van
  Aert, R. C.~M., and van Assen, M. A. L.~M. (2016).
\newblock Degrees of freedom in planning, running, analyzing, and reporting
  psychological studies: A checklist to avoid $p$-hacking.
\newblock {\em Frontiers in Psychology}, 7:1832.
\newblock \doi{10.3389/fpsyg.2016.01832}.

\bibitem[Wright and Ziegler, 2017]{ranger2017}
Wright, M.~N. and Ziegler, A. (2017).
\newblock {ranger}: A fast implementation of random forests for high
  dimensional data in {C++} and {R}.
\newblock {\em Journal of Statistical Software}, 77(1):1--17.
\newblock \doi{10.18637/jss.v077.i01}.

\bibitem[Wynants et~al., 2020]{Wynants2020}
Wynants, L., Calster, B.~V., Collins, G.~S., Riley, R.~D., Heinze, G., Schuit,
  E., Bonten, M. M.~J., Dahly, D.~L., Damen, J.~A., Debray, T. P.~A., de~Jong,
  V. M.~T., Vos, M.~D., Dhiman, P., Haller, M.~C., Harhay, M.~O., Henckaerts,
  L., Heus, P., Kammer, M., et~al. (2020).
\newblock Prediction models for diagnosis and prognosis of covid-19: Systematic
  review and critical appraisal.
\newblock {\em BMJ}, 369:m1328.
\newblock \doi{10.1136/bmj.m1328}.

\bibitem[Yousefi et~al., 2009]{Yousefi2009}
Yousefi, M.~R., Hua, J., Sima, C., and Dougherty, E.~R. (2009).
\newblock Reporting bias when using real data sets to analyze classification
  performance.
\newblock {\em Bioinformatics}, 26(1):68--76.
\newblock \doi{10.1093/bioinformatics/btp605}.

\bibitem[Zou, 2006]{Zou2006}
Zou, H. (2006).
\newblock The adaptive lasso and its oracle properties.
\newblock {\em Journal of the American Statistical Association},
  101(476):1418--1429.
\newblock \doi{10.1198/016214506000000735}.

\bibitem[Zou and Hastie, 2005]{Zou2005}
Zou, H. and Hastie, T. (2005).
\newblock Regularization and variable selection via the elastic net.
\newblock {\em Journal of the Royal Statistical Society: Series B (Statistical
  Methodology)}, 67(2):301--320.
\newblock \doi{10.1111/j.1467-9868.2005.00503.x}.

\end{thebibliography}



\section*{Supplementary material}

# Pitfalls and Potentials in Simulation Studies
## Online Supplement

**Samuel Pawel**[*†], **Lucas Kook**[*], **Kelly Reeve**

Epidemiology, Biostatistics and Prevention Institute

Center for Reproducible Science

University of Zurich, Hirschengraben 84, CH-8001 Zurich

{samuel.pawel, lucasheinrich.kook, kelly.reeve}@uzh.ch

Here, we describe the outcomes of the pre-registered simulations. Overall, the performance of AINET was virtually identical to elastic net regression. The adaptive penalization weights of AINET do not seem to make a difference for the data generating mechanism considered in our simulations. Moreover, since the data were generated under a process equivalent to a logistic regression model, it is no surprise that for reasonably large sample sizes, logistic regression also performed the best. The only exception are conditions with small sample size and low number of events per variable. Here, AINET and elastic net led to more stable and better calibrated predictions than logistic regression. The random forest was outperformed by AINET in most simulation conditions, with exception of very small sample size and prevalence, as well as when a high correlation between covariates was present. Finally, the performance of the adaptive elastic net was generally worse compared to AINET and elastic net. In the following, we summarize the results for each estimand.

## 1 Brier score (primary estimand)

Figure 1 shows the differences in mean Brier score between AINET and the other methods stratified by simulation conditions. We see that there is hardly any difference between AINET and the elastic net (EN) across all simulation conditions meaning that predictive performance of both methods seems to be very similar in the investigated scenarios.

The random forest (RF) shows better predictive performance than AINET in conditions with very low sample size ($n = 100$) and prevalence (prev $= 0.01$). For increasing sample size and prevalence, the performance of AINET seems to become more similar or improve over RF when the correlation of the covariates is not too large ($\rho \leq 0.6$) especially for low events per variable (EPV $\leq 1$). For highly correlated covariates ($\rho = 0.95$), the performance of AINET is similar or worse across most simulation conditions.

Logistic regression (GLM) showed better predictive performance compared to AINET in most simulation conditions. An exception are the conditions with small sample size ($n = 100$), medium to large prevalence (prev $\geq 0.05$) and low events per variable (EPV $\leq 1$), where AINET performed better than GLM.

---

[*]Contributed equally.

[†]Corresponding author: samuel.pawel@uzh.ch

The adaptive elastic net (AEN) method performed worse than AINET in almost all simulation conditions. Only in conditions with very large sample size ($n = 5000$), very small prevalence (prev = 0.01), and high events per variable (EPV = 20), AEN showed predictive performance on par with AINET.

## 2 Scaled Brier score (secondary estimand)

Figure 2 shows the differences in scaled Brier score between AINET and the other methods stratified by simulation conditions. The scaled Brier score is useful to compare the actual values of Brier scores across conditions with different prevalence, but not so much to compare Brier scores of different methods within a simulation condition with fixed prevalence.

We see that for most conditions the plots look like a flipped version of the original Brier scores from Figure 1. Therefore, conclusions are mostly the same. For very small sample sizes coupled with low prevalence and low events per variable (the topleft plots), the scaled Brier score indicates superiority of AINET over RF and GLM, which is opposite the conclusion based on the raw Brier score. We advise to interpret these conditions cautiously since the prevalence prediction which is used for scaling is based on the much larger test data set.

## 3 Log-score (secondary estimand)

Figure 3 shows the differences in log-score between AINET and the other methods stratified by simulation conditions. We see that in certain conditions, the error bars of certain methods are much larger. This is due to the log-score's sensitivity to extreme predictions, which often happen under the RF (and sometimes under the GLM). Despite the larger variability of the log-score, conclusion regarding the comparison between AINET and the other methods are largely the same as under the Brier score.

## 4 Area under the curve (secondary estimand)

Figure 4 shows the differences in area under the curve (AUC) between AINET and the other methods stratified by simulation conditions. As with the other estimands, AINET shows virtually identical performance as EN regression across all simulation conditions. AINET seems to outperform RF across most simulation conditions, with the exception of a conditions with low sample size ($n = 100$), medium prevalence (prev = 0.05), and low events per variable (EPV $\leq$ 1). GLM, typically outperforms AINET conditions with small to medium sample size ($n \leq 500$), and also in conditions with larger sample size when the events per variable is normal to high (EPV $\geq$ 10) and the prevalence is small (prev = 0.01). Finally, the AEN is worse with respect to AUC than AINET across all simulation conditions.

## 5 Calibration slope (secondary estimand)

Figure 5 shows boxplots of calibration slopes stratified by simulation condition and method. For each condition the percentage of simulations where no estimate could be obtained is indicated. This usually happened because of extreme (close to zero or one) predictions, or non-convergence of the method itself.

We caution against interpretation of the random forest (RF) calibration slopes because this method often resulted in predicted probabilities of zero or one, so that a calibration slope could not be fitted.

We see that logistic regression (GLM) shows on average optimal calibration slopes in most simulation condition. In cases where it is off one, its calibration slopes are usually too small indicating overoptimistic predictions. In general, worse calibration slopes are obtained for lower event per variable (EPV).

The penalized methods (AINET, EN, AEN) show a more stable behavior, and on average larger calibration slopes than GLM. This is likely confounded by the simulation conditions in which no GLM calibration slope can be estimated, but estimation of the penalized methods' calibration slope is still possible. Among the penalized method's AINET and EN shows relatively similar calibration slopes whereas the AEN shows worse calibration slopes that are more off the value of one.

## 6 Calibration in the large (secondary estimand)

Figure 6 shows boxplots of calibration in the large estimates stratified by simulation condition and method. For each condition also the percentage of simulations where no estimate could be obtained is indicated. This usually happened because of extreme (close to zero or one) predictions.

We see that the number of simulations with non-estimable calibration is substantially larger when the sample size is small, whereas it decreases for larger sample sizes. An exception is the RF where the number of non-estimable calibrations stays high across most conditions.

While all methods seem to be marginally well calibrated, the penalized methods (AINET, EN, and AEN) show lower numbers of simulations with non-estimable calibration compared to GLM, especially for low to medium sample sizes and low events per variables.

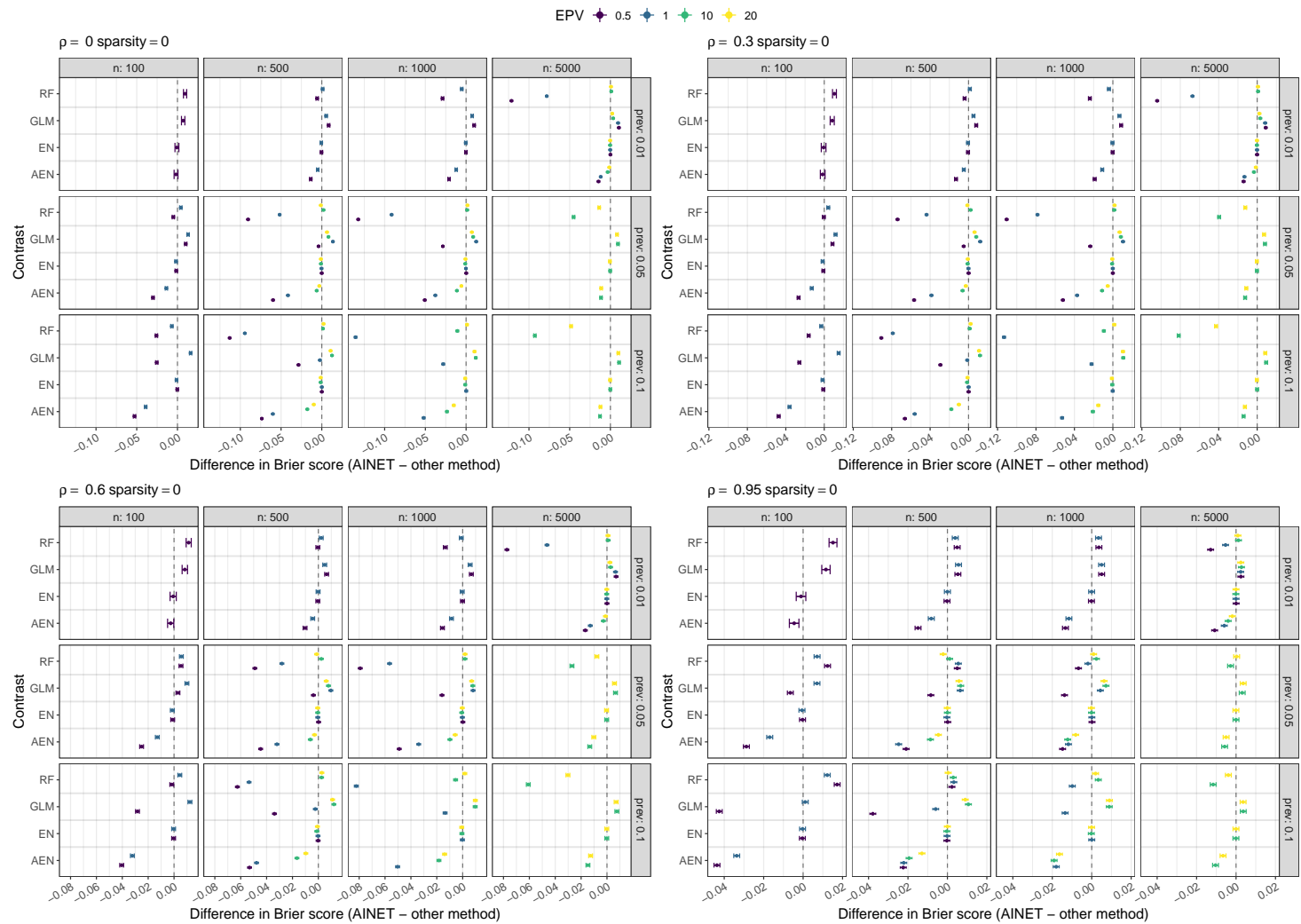

**Figure 1:** Tie-fighter plot for the difference in Brier score between any method on the $y$-axis and AINET. The 95% confidence intervals are adjusted per simulation condition using the single-step method. Lower values indicate better performance of AINET.

4

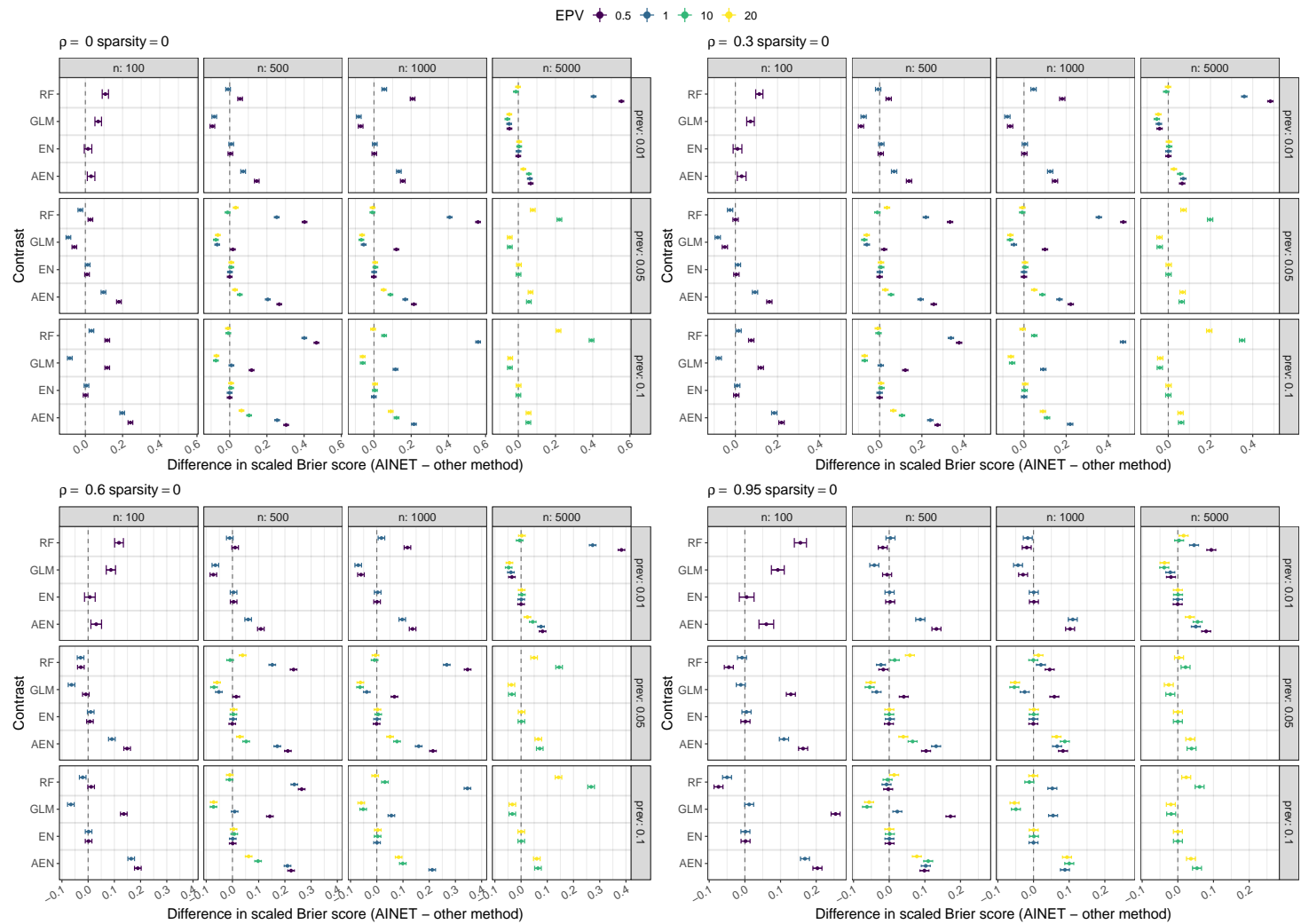

**Figure 2:** Tie-fighter plot for the difference in scaled Brier score between any method on the y-axis and AINET. The 95% confidence intervals are adjusted per simulation condition using the single-step method. Larger values indicate better performance of AINET.

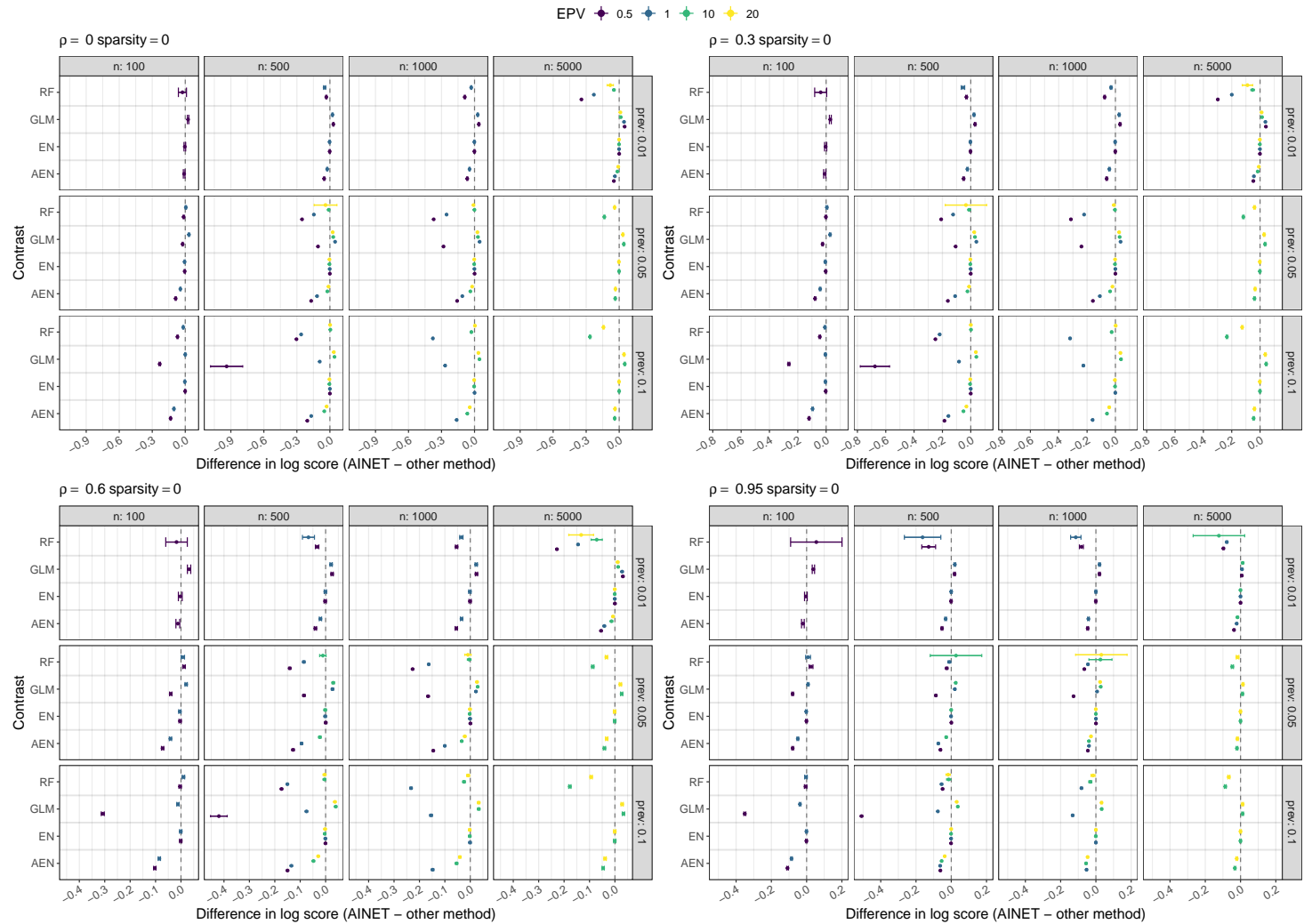

**Figure 3:** Tie-fighter plot for the difference in log-score between any method on the $y$-axis and AINET. The 95% confidence intervals are adjusted per simulation condition using the single-step method. Lower values indicate better performance of AINET.

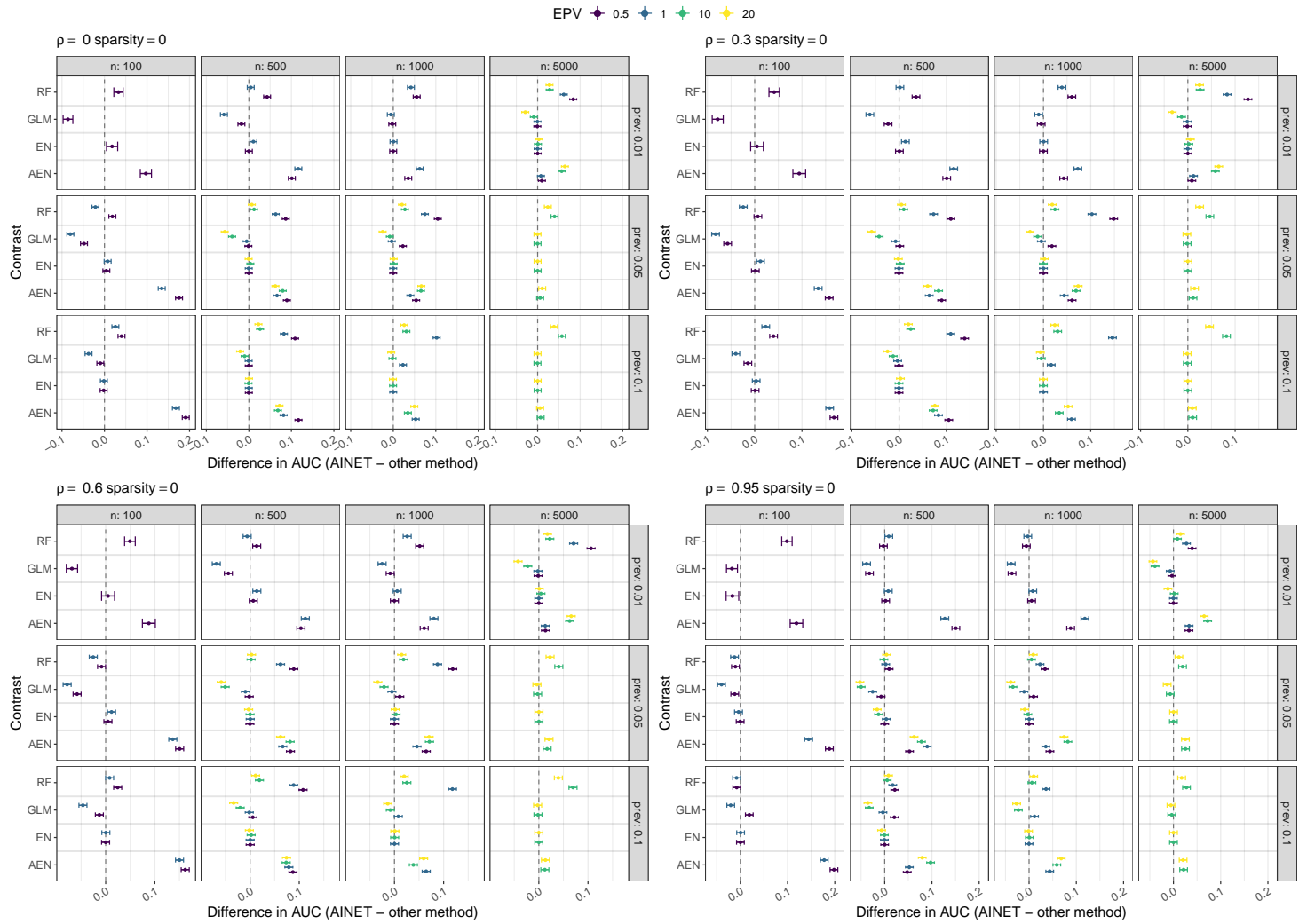

**Figure 4:** Tie-fighter plot for the difference in AUC between any method on the $y$-axis and AINET. The 95% confidence intervals are adjusted per simulation condition using the single-step method. Higher values indicate better performance of AINET.

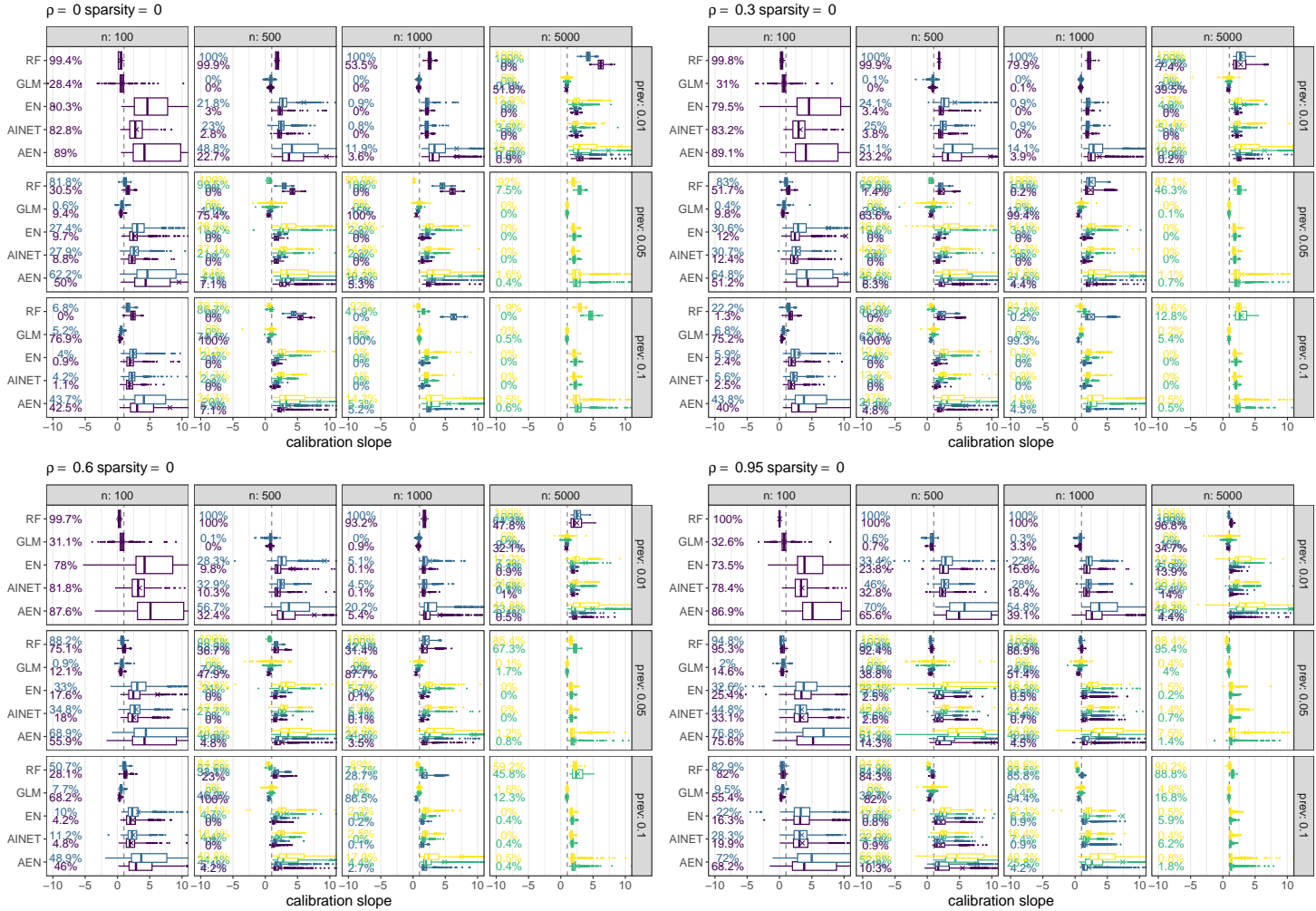

**Figure 5:** Boxplots of calibration slopes stratified by method and simulation conditions. Mean calibration slope is indicated by a cross. A value of one indicates optimal calibration. Percentage of simulations where calibration slope could not be estimated (due to extreme predictions or complete separation) are also indicated.

8

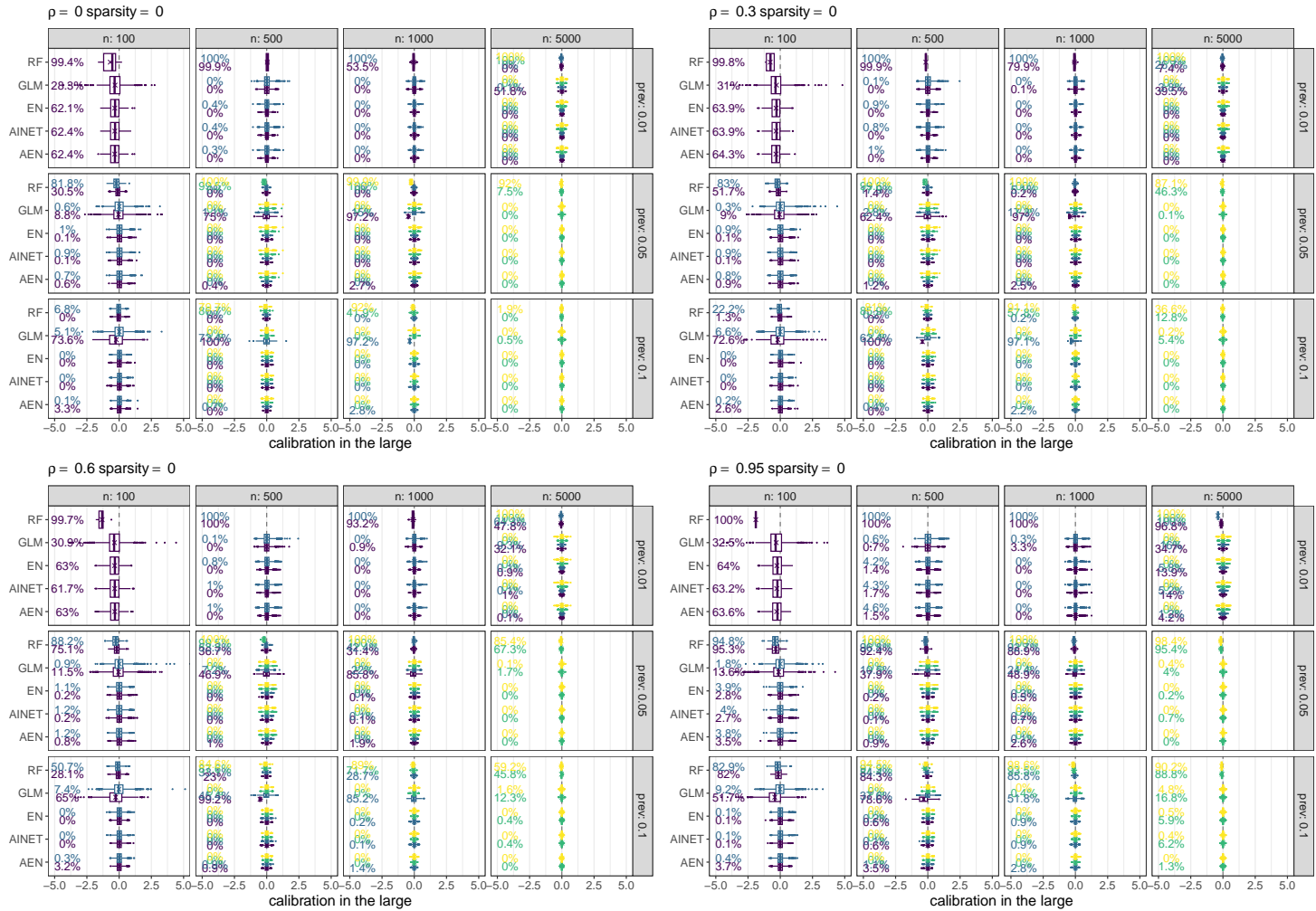

**Figure 6:** Boxplots of calibration in the large stratified by method and simulation conditions. Mean calibration in the large is indicated by a cross. A value of zero indicates optimal calibration in the large. Percentage of simulations where calibration in the large could not be estimated (due to extreme predictions or complete separation) are also indicated.

9



\end{document}